\documentclass[twocolumn,PRA,preprintnumbers,superscriptaddress,amsmath]{revtex4}
\usepackage[latin9]{inputenc}
\usepackage{xcolor}
\usepackage{amsmath}
\usepackage{amssymb}
\usepackage{graphicx}
\usepackage{esint}
\PassOptionsToPackage{normalem}{ulem}
\usepackage{ulem}

\makeatletter

\providecolor{lyxadded}{rgb}{0,0,1}
\providecolor{lyxdeleted}{rgb}{1,0,0}
\@ifundefined{textcolor}{}
{%
 \definecolor{BLACK}{gray}{0}
 \definecolor{WHITE}{gray}{1}
 \definecolor{RED}{rgb}{1,0,0}
 \definecolor{GREEN}{rgb}{0,1,0}
 \definecolor{BLUE}{rgb}{0,0,1}
 \definecolor{CYAN}{cmyk}{1,0,0,0}
 \definecolor{MAGENTA}{cmyk}{0,1,0,0}
 \definecolor{YELLOW}{cmyk}{0,0,1,0}
}

\makeatother

\begin{document}

\title{Magnon dark modes and gradient memory}

\author{Xufeng Zhang}
\thanks{These authors contributed equally to
this work.}
\address{Department of Electrical Engineering, Yale University, New Haven,
Connecticut 06511, USA}

\author{Chang-Ling Zou}\thanks{These authors contributed equally to
this work.}
\address{Department of Electrical Engineering, Yale University, New Haven,
Connecticut 06511, USA}
\address{Department of
Applied Physics, Yale University, New Haven,
Connecticut 06511, USA}
\address{Key Lab of Quantum Information, University of Science and Technology
of China, CAS, Hefei, Anhui, 230026, China}

\author{Na Zhu}
\address{Department of Electrical Engineering, Yale University, New Haven,
Connecticut 06511, USA}

\author{Florian Marquardt}
\address{Institute for
Theoretical Physics II, University of
Erlangen-Nuremberg Staudtstr. 7, 91058 Erlangen,
Germany}
\address{Max Planck Institute for the Science of Light, G\"{u}nther-Scharowsky-Stra\ss e1/Bau 24, 91058 Erlangen, Germany}

\author{Liang Jiang}
\address{Department of Applied Physics, Yale University, New Haven, Connecticut
06511, USA}

\author{Hong X. Tang}
\address{Department of Electrical Engineering, Yale University, New Haven,
Connecticut 06511, USA}

\maketitle \textbf{Extensive efforts have been
expended in developing hybrid quantum systems to
overcome the short coherence time of
superconducting circuits by introducing the
naturally long-lived spin degree of freedom.
Among all the possible materials, single-crystal
yttrium iron garnet has shown up very recently as
a promising candidate for hybrid systems, and
various highly coherent interactions, including
strong and even ultra-strong coupling, have been
demonstrated. One distinct advantage of these
systems is that the spins are in the form of
well-defined magnon modes, which allows flexible
and precise tuning. Here we demonstrate that by
dissipation engineering, a non-Markovian
interaction dynamics between the magnon and the
microwave cavity photon can be achieved. Such a
process enables us to build a magnon gradient
memory to store information in the magnon dark
modes, which decouple from the microwave cavity
and thus preserve a long lifetime. Our findings
provide a promising approach for developing
long-lifetime, multimode quantum memories.}

Hybrid systems provide a promising solution for
coherent information storage by combining the
long coherence time of spin ensembles with the
power of superconducting circuits
\cite{Wallquist2009, Xiang2013}. Researches
utilizing ensembles ranging from cold atomic
gases \cite{Verdu2009PRL_atom} and magnetic
molecules \cite{Eddins2014PRL_Molecule} to
rare-earth-ion-doped crystals
\cite{SchusterPRL2010_Ruby, ProbstPRL2013_ErYSO,
Tkalcec2014_PRB} and negatively-charged nitrogen
vacancy (NV$^{-}$) centres in diamond
\cite{ZhuNature2011, Kubo2011PRL_qubit,
AmsussPRL2011, Marcos2010PRL, Ranjan2013PRL} have
been reported recently. But not until very
recently did people start to investigate the
possibility of hybridizing yttrium iron garnet
(YIG, Y$_{3}$Fe$_{5}$O$_{12}$), a ferrimagnetic
insulator, with microwave cavities. In the YIG
crystal, the coherent photon-spin ensemble
interaction is greatly enhanced by the high spin
density and can even approach the ultrastrong
coupling regime \cite{Huebl2013PRL_YIG,
Zhang2014_PRL, Tabuchi2014}. Compared with
previous dilute spin ensemble systems, direct
spin-spin interaction leads to collective magnon
excitations that have a range of distinct
advantages, such as low damping rate, uniform
distribution, rich nonlinear dynamics, and
well-defined mode profiles and wave-vectors.
Particularly, each piece of YIG can be treated as
a giant spin, with the flexibility to be
individually manipulated in experiments.

In this work, we strongly couple multiple magnon
modes with a microwave cavity resonance by
placing multiple YIG spheres into a three
dimensional (3D) cavity. With these coherently
coupled magnons, collective effects
\cite{Dicke1954} such as a magnon dark mode
(subradiant mode) that is decoupled from the
environment, as well as enhanced interaction
between a magnon bright mode (superradiant mode)
and a microwave mode are demonstrated. The
spectrum of the magnons can be further adjusted
by controlling the local magnetic field of each
YIG sphere, which allows us to engineer the
dissipation and tailor the dynamics of the
microwave photon at will. With this method, we
applied a magnetic field gradient, which induced
the periodic evolution of the magnons between
their temporal dark and bright modes, leading to
a non-Markovian dynamics of the cavity energy
that shows non-exponential decay and revival
\cite{Zou2013_PRA}. By optimizing the coupling
condition \cite{Afzelius2010_PRA}, good
efficiency is obtained and our theory analysis
indicates that a unity efficiency is achievable.
Our experiment is performed at room temperature,
demonstrating the coherent, long-lifetime,
broadband and multimode gradient memory effect.
Since the principle is entirely based on linear
interactions, such a magnon memory can be readily
scaled down to the quantum regime at millikelvin
temperatures, providing a new approach besides
the existing schemes for dilute spin ensembles
\cite{Hau1999_Nature, Anderson1955_JAP,
Alexander2006_PRL, Hosseini2011_NatComm,
Hedges2010_Nature, Kraus2006_PRA, Arnold1991_OL,
DeRiedmatten2008_Nature}. This allows the
realization of hybrid quantum memories that do
not suffer from common problems such as
inhomogeneous broadening and so on.

\begin{figure}[t]
\includegraphics[width=8cm]{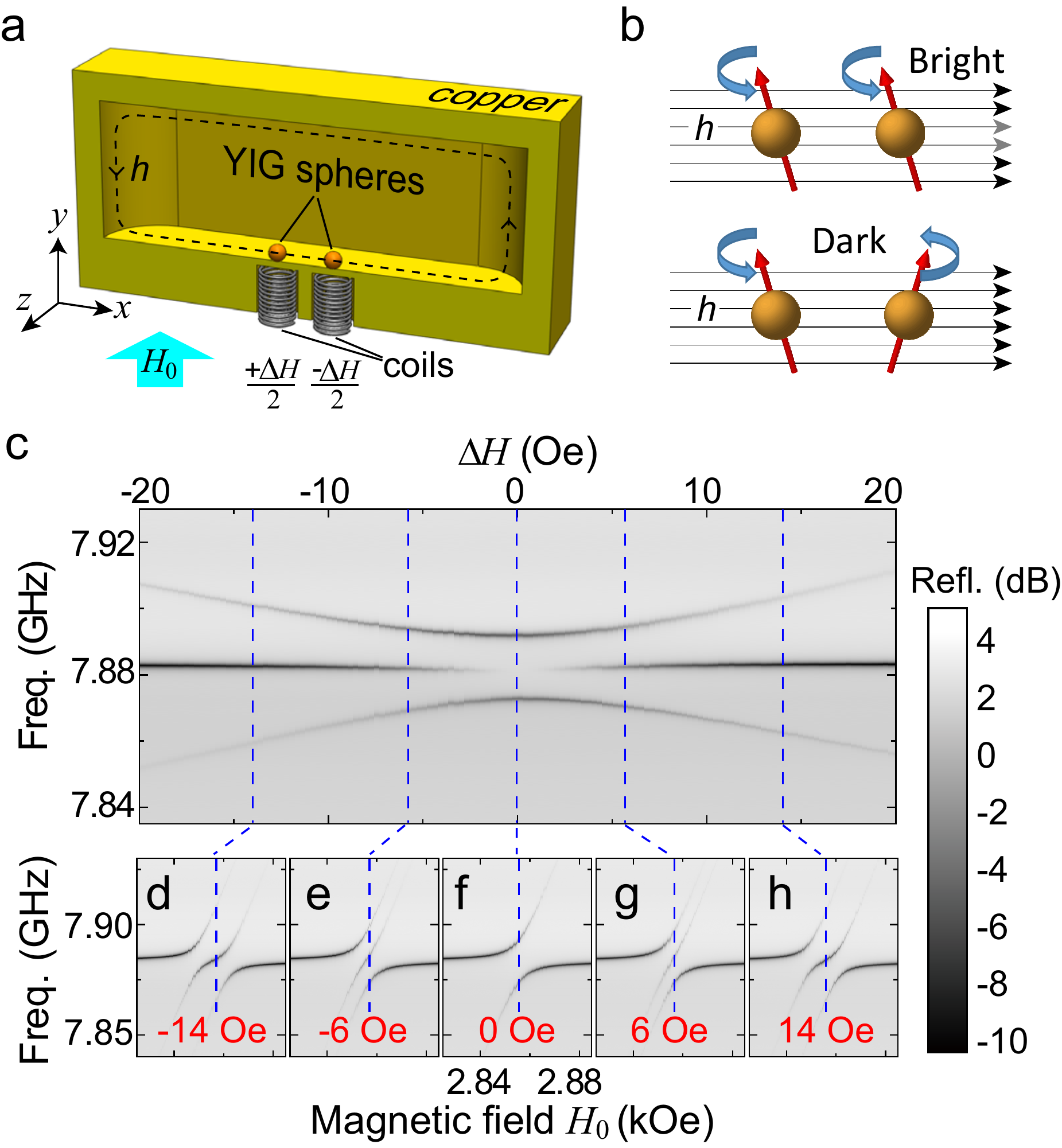}
\caption{\textbf{Magnon dark mode of two YIG
spheres.} \textbf{a,} Device schematic including
half of the copper cavity, two YIG spheres, and
two small coils. $H_0$: external bias magnetic
field; $\Delta H$: magnetic field gradient
generated by the small coils; $h$: magnetic field
of the microwave resonance (TE$_{110}$ mode) in
the copper cavity. \textbf{b,} Conceptual
illustration of the magnon bright and dark modes:
the bright mode is coupled to while the dark mode
is isolated from the microwave magnetic field
$h$. \textbf{c,} Reflection spectra as a function
of $\Delta H$ under an external bias magnetic
field $H_0=2858$ Oe. \textbf{d-h,} Reflection
spectra as a function of bias magnetic field
$H_0$ at a fixed magnetic field gradient $\Delta
H=-14,-6,0,6,14$ Oe, respectively. Each dashed
line in \textbf{c} corresponds to the dashed line
in each of the spectral maps in \textbf{d-h}.}

\label{fig:1}
\end{figure}

As a general situation, consider $N$ identical YIG spheres loaded
in a copper microwave cavity. The linear coupling between the uniform
magnon modes in the YIG spheres and the cavity TE$_{110}$ mode can
be described by the Hamiltonian
\begin{equation}
\mathcal{\mathcal{\hat{H}}/\hbar}=\omega_{a}\hat{a}^{\dagger}\hat{a}+\sum_{j=1}^{N}(\omega_{j}\hat{m}_{j}^{\dagger}\hat{m}_{j}+g_{j}\hat{a}^{\dagger}\hat{m}_{j}+g_{j}^{*}\hat{a}\hat{m}_{j}^{\dagger}),\label{eq:Hamiltonian}
\end{equation}

\noindent where $\hat{a},\ \hat{m}_{j}$ are the
bosonic operators associated with the photon and
the uniform magnon modes, respectively, $g_{j}$
is the coupling strength between the microwave
mode and the magnon mode in the $j$-th YIG sphere
with $j$ ranging from 1 to $N$ being the sphere
index, $\omega_{a}$ is the resonance frequency of
the TE$_{110}$ mode of the copper cavity with the
YIG spheres loaded but at zero bias field,
$\omega_{j}=\omega_{a}+[j-(N+1)/2]\Delta\omega_{m}$
are the evenly distributed magnon resonance
frequencies, and $\Delta\omega_{m}$ is their
frequency interval. Such a frequency distribution
is obtained by biasing each YIG sphere at a
magnetic field $H_{j}=H_{0}+[j-(N+1)/2]\Delta H$,
where $H_{0}=\omega_{a}/\gamma$ is the common
external magnetic field that brings a YIG sphere
on resonance with the microwave cavity TE$_{110}$
mode, $\gamma/2\pi=2.8$ MHz/Oe is the
gyromagnetic ratio, and $\Delta
H=\Delta\omega_{m}/\gamma$ is the magnetic field
difference between neighboring YIG spheres that
is provided by the fine tuning of the small coils
underneath each individual YIG sphere
(Fig.\,\ref{fig:1}a).

We first study the coherence of a simple system
with two YIG spheres ($N=2$). As a result of the
mode hybridization for nonzero $g_{1,2}$, there
exist three resonances in the cavity reflection
spectrum when the two magnon modes are near
resonance with the cavity TE$_{110}$ mode but not
on resonance, which agrees well with our
experimental observation (Fig.\,\ref{fig:1}c). By
fixing $\Delta H$ and sweeping the external
magnetic field around $H_{0}$, two avoided
crossings are observed in the reflection spectra
(Fig.\,\ref{fig:1}d), indicating the strong
coupling between the cavity mode and the two
magnon modes. When the frequencies of the two
magnon modes ($\omega_{1,2}$) are brought closer
by tuning $\Delta H$, the absorption of the
middle resonance becomes weaker, which eventually
vanishes in the spectrum when the two magnon
modes are brought simultaneously on resonance
with the cavity mode
($\omega_{1}=\omega_{2}=\omega_{a}$) by turning
off the gradient ($\Delta H=0$), and such a
transition is evident in Figs.\,\ref{fig:1}c-h.
The remaining two resonances emerge from the
hybridization of the bright (superradiant) magnon
mode with the cavity photon mode, while the
resonance that has disappeared from the spectrum
is the dark (subradiant) magnon mode, as it
decouples from the microwave cavity.

The concepts of the bright and dark modes are
illustrated schematically in Fig.\,\ref{fig:1}b.
When the detuning field $\Delta H=0$, the two
magnon modes are on resonance
$\omega_{1}=\omega_{2}$. Since, in addition, we
have $g_{1}=g_{2}$, the bright mode is the
superposition of the two magnon modes that
precess in phase $\hat{B}
=\frac{1}{\sqrt{2}}\left(\hat{m}_{1} +\hat{m}_{2}
\right)$, while the dark mode is the
superposition of the two magnon modes that
precess out of phase $\hat{D}
=\frac{1}{\sqrt{2}}\left(\hat{m}_{1} -\hat{m}_{2}
\right)$. For the bright mode, the coherent
interactions between the magnons and photons are
collectively enhanced, leading to an enhanced
coupling strength
$g_{\mathrm{B}}=\sqrt{2}g_{1,2}$, which is
verified in the avoided crossing spectrum where
the splitting for $\Delta H=0$ ($2g/2\pi=18.88$
MHz, Fig.\,\ref{fig:1}f) is $\sqrt{2}$ times
larger than that for $\Delta H=\pm14\
\mathrm{Oe}$ ($2g/2\pi=13.42$ MHz,
Figs.\,\ref{fig:1}d\&h). For the dark mode, the
coupling of the two magnon modes with the cavity
photons cancel each other, resulting in a
vanishing coupling strength $g_{\mathrm{D}}=0$.

\begin{figure*}[t]
\centering{}
\includegraphics[width=17cm]{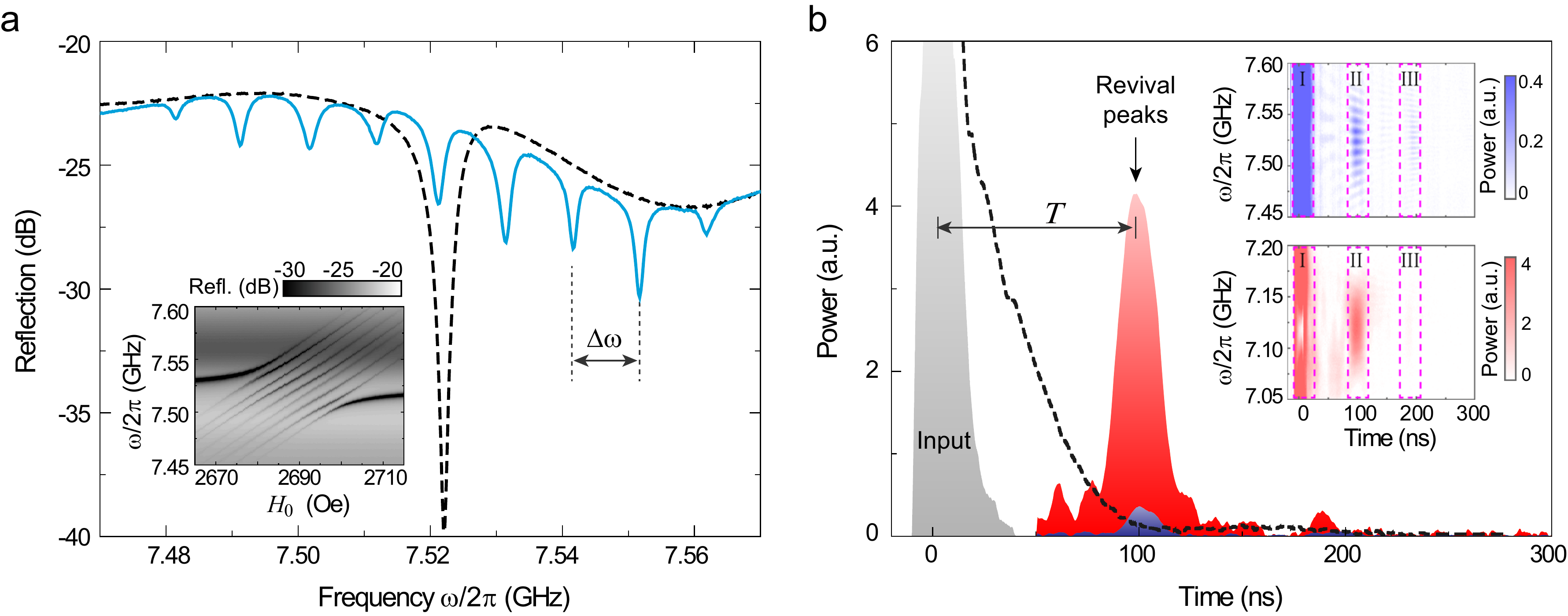}
\caption{\textbf{The magnon gradient memory.}
\textbf{a,} Reflection spectra of the microwave
cavity with eight YIG spheres at a bias field
($H_{0}$) of $2687$ Oe (solid blue line) and $0$
Oe (dashed black line), respectively. The
frequency gradient of the magnon modes is tuned
to be $\Delta\omega=2\pi\times10$ MHz. Inset: the
reflection frequency spectra of the MGM at
various bias magnetic fields. \textbf{b,} The
dynamics of the MGM output for a 15 ns pulsed
microwave excitation (gray) at the under-coupling
(blue) and critical-coupling (red) conditions
with a frequency gradient of
$\Delta\omega=2\pi\times10$ MHz. Dashed black
curve shows the exponential decay when the
magnons are largely detuned. Operation condition
for the blue curve: $\omega=2\pi\times7.522$ GHz,
$H_{0}=2687$ Oe; black: $\omega=2\pi\times7.522$
GHz, $H_{0}=0$ Oe; red: $\omega=2\pi\times7.120$
GHz, $H_{0}=2544$ Oe. The input frequency for
each case is adjusted accordingly with the cavity
resonance shift induced by the change of the
external coupling. Inset: pulse retrieval at
various frequencies for the under-coupling (top)
and critical-coupling (bottom) condition,
respectively. Zone I: reflection at the time of
input due to the under-coupling condition of the
external coupling; Zone II: first revival peaks;
Zone III: second revival peaks.}

\label{fig:2}
\end{figure*}

While the magnon bright mode can be used for
information conversion with microwave photons,
the dark mode is an ideal candidate for
information storage because it decouples from the
cavity and therefore has a very long lifetime
\cite{Dong2012_Science}. A dark mode memory can
be constructed if fast conversion between the
bright and the dark mode is available, which in
principle can be realized by rapidly tuning the
magnetic bias field. However, the slow response
of the local inductive coils prohibits the
experimental realization of such an approach.
Alternatively, we consider here a magnon gradient
memory (MGM) using magnon \emph{temporal dark
modes} which eliminates the need of fast magnetic
control. Such temporal dark modes are created by
applying a magnetic field gradient ($\Delta
H\neq0$) to uniformly detune the magnon modes
around the cavity resonance, so unlike the steady
bright and dark modes discussed above, they are
not the eigenmodes of the Hamiltonian. Instead,
the non-Markovian dynamics of such a coupled
system leads to a temporally evolving
inter-conversion between the temporal bright and
dark modes, which provides a good solution for
fast memory operations without adding extra
controls.

For a device consisting of $N$ YIG spheres, a
frequency spectrum with $N+1$ resonances spaced
at an interval of $\Delta\omega$
($\approx\Delta\omega_{m}$ for large $N$) can be
obtained as a result of the coupling between the
cavity mode and the uniformly distributed magnon
modes. To achieve a memory with high efficiency,
it is also important to ensure uniform coupling
strengths for all the YIG spheres ($g_{j}=g_{0}$,
see Supplementary Information for details). The
collective magnon modes from the $N$ YIG spheres
can be expressed as one temporal bright mode,
expressed as
\begin{equation}
\hat{B}=\frac{1}{\sqrt{N}}\sum_{j=1}^{N}\hat{m}_{j},\label{eq:bright}
\end{equation}

\noindent and $N-1$ temporal dark modes, with the
$n$-th dark mode expressed as

\begin{equation}
\hat{D}_{n}=\frac{1}{\sqrt{N}}\sum_{j=1}^{N}\hat{m}_{j}e^{2\pi
i(j-\frac{N+1}{2})\cdot n/N},\label{eq:dark}
\end{equation}

\noindent which do not couple to the cavity. Due
to the uniformly-spaced detuning, the magnon
modes described by Eqs.\,(\ref{eq:bright}) and
(\ref{eq:dark}) will convert from one to another
with a time interval $T/N$, where
$T=2\pi/\Delta\omega$ is the evolution period of
the collective magnon modes. During the storing
process, the cavity photons are first converted
to the temporal bright magnon mode, which is then
successively converted to the temporal dark
magnon modes that prevent the radiation loss.
After one period, the system evolves back to the
bright mode, and the information retrieves from
the magnons and converts back to microwave
photons. Both storing and retrieving processes
are accelerated due to the superradiance effect.
In such a configuration, no external control,
such as refocusing pulse or gradient inversion,
is required to retrieve the stored information.
The performance of such an MGM depends on the
number of YIG spheres. A larger $N$ helps to
improve the pulse re-construction and suppress
the off-peak ripples (Supplementary Information).

In our experiments, we demonstrate such an MGM
using eight YIG spheres ($N=8$). The large
tunability of the magnon allows us to engineer
the cavity dissipation with great flexibility and
obtain a reflection spectrum with a total of
$N+1=9$ uniformly distributed hybrid modes, as
shown in Fig.\,\ref{fig:2}a (solid blue line),
where $\Delta\omega/2\pi=10$ MHz. Note that, in
contrast to the stationary dark mode, the
temporal dark modes can be detected in the
spectrum since they convert back to the bright
mode periodically. In other words, each of the 9
resonance lines in the steady state spectrum
contains some component from the temporal bright
mode which can be detected. When a 15-ns-long
pulsed microwave signal at a frequency
$\omega=\omega_{a}=7.52$ GHz is injected into the
cavity with an external bias magnetic field of
2687 Oe, it couples to the magnon bright mode and
then converts to the magnon dark modes. The
retrieval of the stored pulse takes place after a
pre-programmed time $T=2\pi/\Delta\omega=100$ ns
(blue peak in Fig.\,\ref{fig:2}b), in sharp
contrast with the exponential decay when the
magnons are strongly detuned by turning off the
external bias magnetic field $H_{0}$ (dashed
black line in Fig.\,\ref{fig:2}b). While the MGM
works without the requirement of any
time-dependent control, extensions of the scheme
would be possible when that control becomes
available. For instance, on-demand recall can be
achieved via dynamic control by rapidly turning
off and on the magnetic field gradient. Moreover,
further dissipation engineering by controlling
the field gradient would allow even more
complicated manipulation of the output pulses,
such as sequence reversal, pulse splitting, and
so on. However, a fast magnetic field tuning at
this time scale with reasonably large amplitude
(around $50\ \mathrm{Oe}$) is difficult to
achieve and so it falls beyond the scope of this
work and is left for future study.

The performance of the MGM can be significantly
improved by optimizing the external coupling
condition. Blue curves in Figs.\,\ref{fig:2}a \&
b correspond to the condition of under-coupling
and therefore the retrieved pulse is very weak. A
drastic boost of the revival pulse is obtained
(red peak in Fig.\,\ref{fig:2}b) by adjusting the
coaxial probe to meet the critical coupling
(impedance matching) condition:
$\kappa_{a,1}=\kappa_{a,0}+\pi|g|^{2}/\Delta\omega$,
where $\kappa_{a,0}$ is the amplitude damping
rate of the cavity resonance and $\kappa_{a,1}$
is the coupling rate from the coaxial probe. The
critical coupling condition allows a complete
energy conversion into the magnon bright mode
(see Supplementary Information), and it also
results in a maximized signal retrieval back to
the microwave photon. Therefore, the first
revival peak will contain all the stored
information and there will be no successive
revival peaks as observed in the under-coupling
situation. Such analysis matches our measurement
results shown in the insets of
Fig.\,\ref{fig:2}b: the second revival peaks
(zone III) are relatively strong compared with
the first revival peaks (zone II) for the
under-coupling situation (top inset), while for
the critical-coupling situation (bottom inset)
the second revival peaks disappear.

One advantage of the MGM is that its operation is
not restricted to a narrow frequency range. A
uniform frequency distribution exists in a wide
frequency range as we vary the bias magnetic
field (Fig.\,\ref{fig:2}a, inset). Accordingly,
broad-band pulse retrieval process can be
obtained in the time-domain measurement.
Generally, the operation bandwidth can be
determined by $N\Delta\omega$, and therefore, the
delay-bandwidth product is set by $N$. In the
experiment, we observed retrieval peaks in a
broad frequency band of 80 MHz at a bias magnetic
field of 2687 Oe (Fig.\,\ref{fig:2}b, insets),
which is 1-2 orders of magnitude larger than the
linewidth of the photon or magnon resonances.
Note that when sweeping the input frequency, the
revival peak intensity periodically varies for
the under-coupling situation (Fig.\,\ref{fig:2}b,
top inset, zone II), which can be attributed to
the periodic variation in the coupling condition,
while such a phenomenon disappears in the
critically coupled situation thanks to the
improved coupling condition.

\begin{figure}[t]
\centering{}\includegraphics[width=8cm]{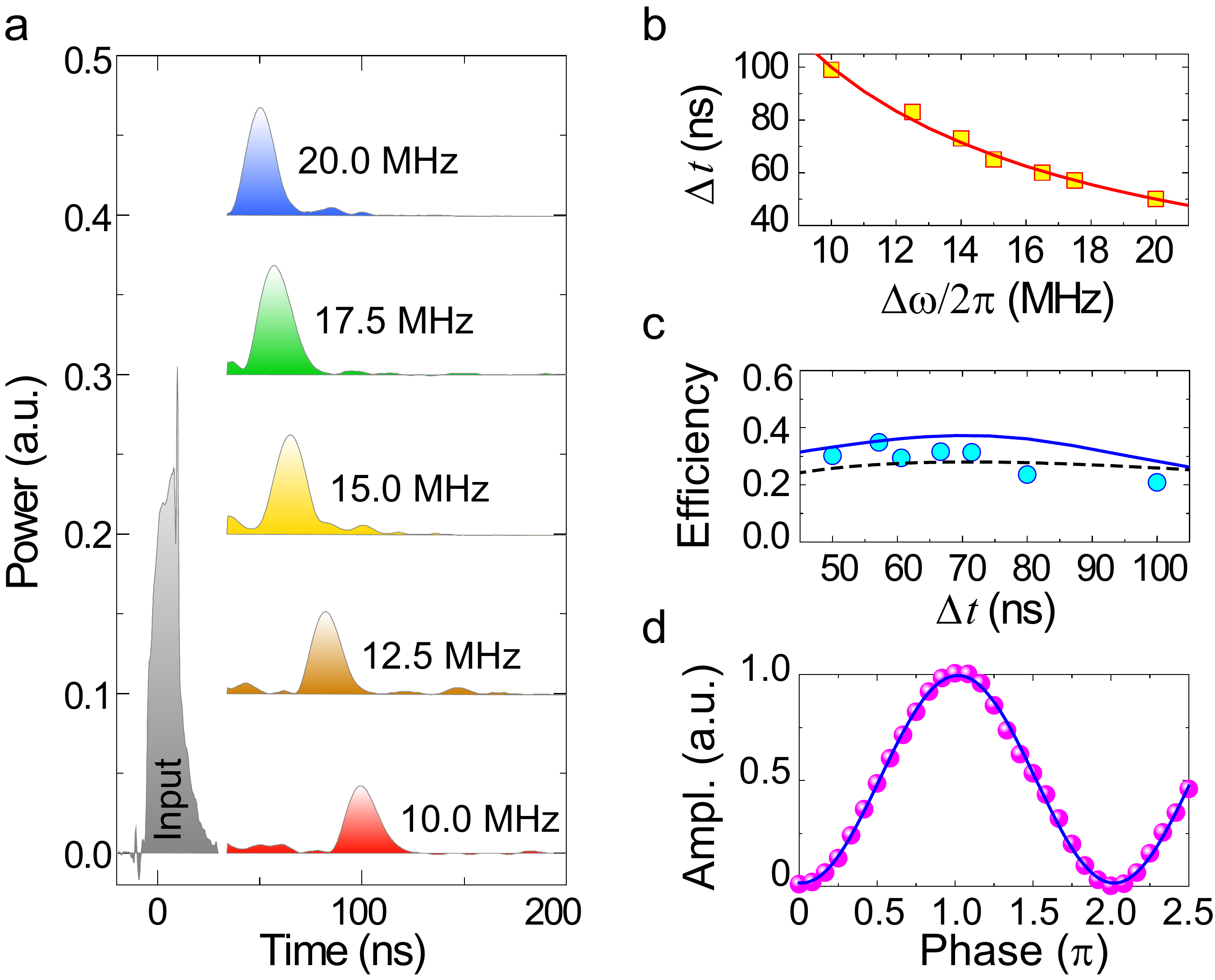}
\caption{\textbf{Characterization of magnon
gradient memory. a,} Measured retrieval pulses
for different frequency gradients
$\Delta\omega/2\pi$. The input pulse is only
shown for $\Delta\omega/2\pi=10$ MHz and remains
the same for the other cases. \textbf{b,}
Extracted storage time as a function of frequency
gradient. The solid line is calculated by
$T=2\pi/\Delta\omega$. \textbf{c,} Retrieval
efficiency as a function of the storage time
obtained from the measurement (circles), the
numerical fitting (solid blue line), and the
calculation using Eq.\,(\ref{eq:zeta}) (dashed
black line), respectively. \textbf{d,} Measured
interference between the retrieved and reference
signals as a function of their phase difference.
The solid line is a sinusoidal fitting.}

\label{fig:3}
\end{figure}

We quantitatively characterized the MGM by
measuring the delay time and the retrieval
efficiency as a function of $\Delta\omega$, and
the result is plotted in Fig.\,\ref{fig:3}a. As
$\Delta\omega$ reduces, the revival time $T$
increases, and the measurement results perfectly
follow the relation $T=2\pi/\Delta\omega$
(Fig.\,\ref{fig:3}b). The MGM retrieval
efficiency is defined as the ratio of the output
pulse energy to the input pulse energy. It has
been extracted as a function of delay time and
plotted in Fig.\,\ref{fig:3}c (blue circles),
showing an efficiency of about 30\% under the
critical coupling condition. The extracted
efficiencies match the numerical simulation
results obtained using the same parameters as in
the measurement (solid blue line). On the other
hand, the efficiency can be approximated using
the asymptotic expression (see Supplementary
Information)
\begin{equation}
\zeta\approx e^{-2\pi/F}\left[1-\frac{1}{t_{p}\kappa_{a,0}\left(1+G\right)}\right]\left(\frac{G}{1+G}\right)^{2},\label{eq:zeta}
\end{equation}
which is plotted in Fig.\,\ref{fig:3}c (dashed
black) for comparison. Equation (\ref{eq:zeta})
shows that the MGM efficiency depends on the
figure of merit $G=\frac{\pi}{2}\frac{C}{F}$ for
an input pulse with a duration of $t_{p}$, where
$F=\frac{\Delta\omega}{2\kappa_{m}}$ is the
finesse of the magnon gradient and
$C=\frac{\left|g\right|^{2}}{\kappa_{m}\kappa_{a,0}}$
is the magnon-photon interaction cooperativity.
Here we assume all the magnon modes in each YIG
sphere have the same amplitude damping rate
($\kappa_{m}$). Therefore by optimizing $C$ and
$F$ we can further improve the MGM performance.
In practice, $F$ can be well controlled to be in
the range of $10^{1}-10^{2}$, and $C$ in the
order of $10^{2}-10^{4}$, and thus
$\zeta\approx1$ is achievable for real
applications.

\noindent
\begin{figure}[t]
\includegraphics[width=7.5cm]{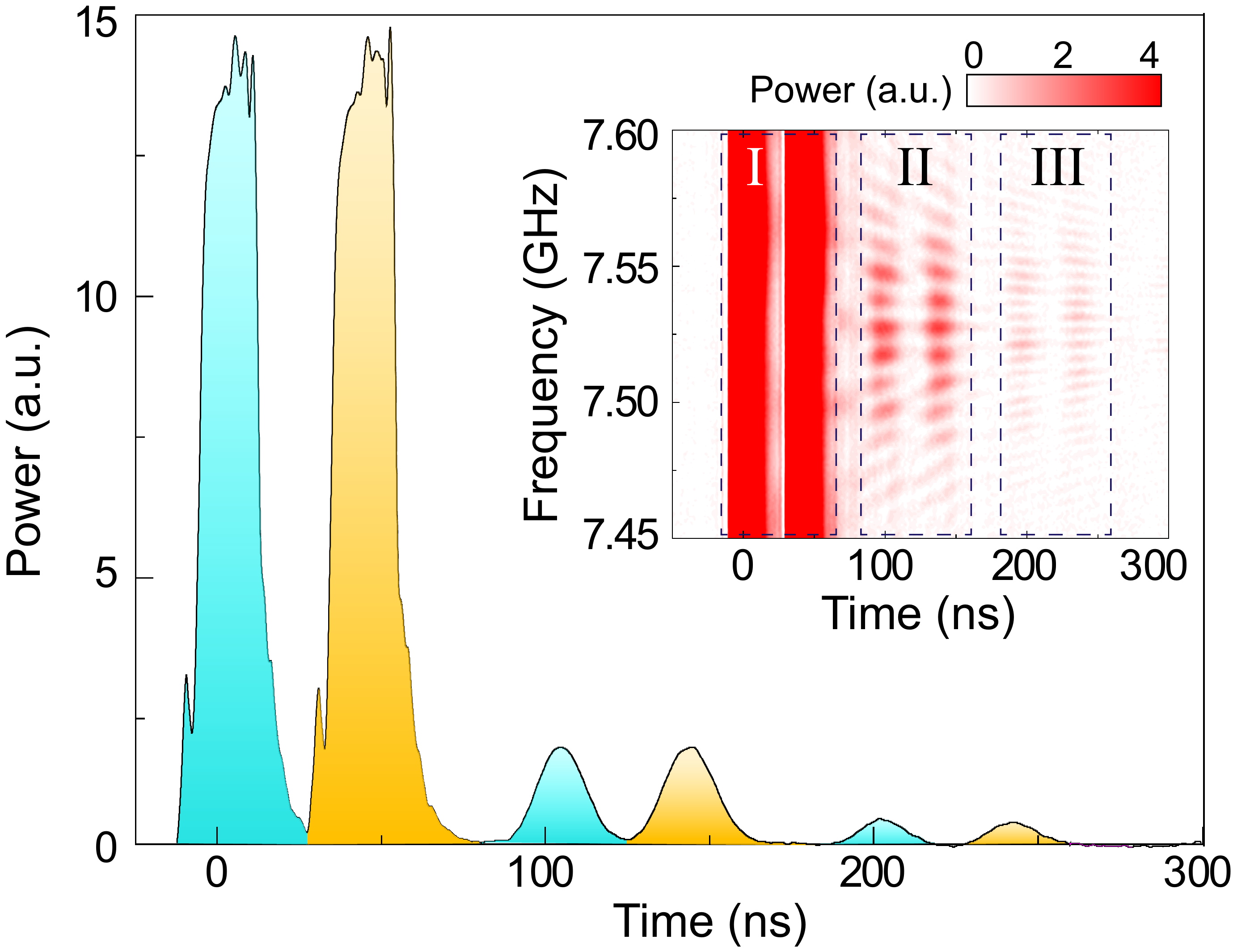} \caption{\textbf{Multi-pulse storage in the MGM.} Retrieved pulses for a
double-pulse excitation. The two pulses are
separated by 40 ns with a 15 ns duration each.
Inset: double-pulse retrieval for various input
frequencies at a bias field of 2687 Oe.}

\label{fig:4}
\end{figure}

In addition to the high achievable efficiency,
such an MGM preserves the signal coherence during
the storage and retrieval process.
Figure\,\ref{fig:3}d plots the measured
interference signal between the retrieved pulse
(after a 100-ns storage time) and the input
microwave signal. The periodic dependence of the
interferometer output on the relative phase
between these two signals indicates they are
still phase coherent. A visibility as high as
$96.9\pm1.5$\% can be extracted, proving the high
coherence preservation feature of our MGM.

Another advantage of the MGM is its multimode
operation capability, which is crucial for
speeding up quantum computation protocols.
Figure\,\ref{fig:4} plots the output of the MGM
when two identical pulses are sent into the
memory. The input pulses are 15 ns in duration
and separated by 40 ns. Two pulses are retrieved
from the memory after the pre-programmed storage
time $T=100$ ns. As there is no echo process
involved during the memory process, the MGM
operates in a first-in-first-out (FIFO) mode. The
multimode operation can be also achieved in a
wide frequency range (Fig.\,\ref{fig:4}, inset).
The occasional pulse distortion at certain
frequencies is a result of the non-ideality of
the system such as the non-identical coupling
strengths and decay rates of different YIG
spheres, or the impact of the high order magnon
modes.

In summary, we demonstrated coherent coupling of
multimode magnon resonances to a single microwave
cavity. The dark magnon mode of two YIG spheres
is observed. We used the temporal magnon dark
mode to establish non-Markovian dynamics in a
system consisting of eight YIG spheres, and
showed its potential in developing quantum
memories, which are broadband, multimode, and do
not require fast field switching. Our
investigation on a novel mechanism of
manipulating magnons and microwave photons paves
the route towards magnon-based hybrid quantum
memories, quantum repeaters as well as quantum
networks \cite{Kimble2008}.

\vbox{}

\noindent \textbf{\large{}{}{}Methods}{\large \par}

\noindent \textbf{Sample preparation.} The
microwave cavities are machined from
high-conductivity copper. The cavity housing two
YIG spheres has a dimension of
$43\times21\times9.6$ mm$^{3}$, with its
TE$_{110}$ resonating at 7.87 GHz, while the one
housing eight YIG spheres has a dimension of
$50\times21\times4.8$ mm$^{3}$, with its
TE$_{110}$ mode resonating at 7.52 GHz. All the
YIG spheres are highly polished and have a
diameter of 0.25 mm, with their $\left\langle
110\right\rangle $ direction aligned parallel to
the external bias magnetic field. The mounting
ceramic rods that hold the YIG spheres are glued
on the cavity wall, close to the the maximum
magnetic field of the cavity resonance to achieve
strong magnon-photon coupling. Eight holes, 6.5
mm apart from each other, are drilled deep into
(but not through) the cavity wall underneath the
YIG spheres, where eight home-made small coils
(200 turns, 6 mm in diameter, made of 32-gauge
copper wires) are inserted with a 1 mm distance
from the YIG spheres for fine tuning of the bias
magnetic fields within a range of $\pm$30 Oe
(Figs.\,1a \& S5a). The whole device assembly is
inserted between the two poles of a water-cooled
electromagnet which provides the strong external
bias magnetic field.

\vbox{}

\noindent \textbf{Microwave characterization.}
The frequency spectra are taken using a vector
network analyzer by measuring the reflection
signal from the coaxial probe which accesses the
cavity mode through a small hole in the cavity
wall. The external coupling can be tuned by
adjusting the position of the coaxial probe. The
time traces are measured using a high-speed
oscilloscope when the continuous-wave (CW) input
microwave signal is modulated by a pulse
generator through a transistor\textendash
transistor logic (TTL) switch which gives a
rectangular pulse shape. To measure the coherence
between the retrieved pulse and the input signal,
the CW input signal is split into two branches
before the pulse modulation, one of which,
serving as the reference, is phase adjusted and
combined with the retrieved signal before sending
into an envelope detector. Then, the visibility
of interference between the retrieved signal and
the reference is measured by varying the
reference phase. More details about the microwave
characterization can be found in the
Supplementary Information.

\vbox{}

\noindent \textbf{Theoretical derivation.} In the
Supplementary Information, details about the
theory derivation and analysis of the dark mode
and the gradient memory are provided.

\vbox{}

\noindent \textbf{Acknowledgments}\\
 This work is supported by DARPA/MTO MESO program. H.X.T. acknowledges
support from a Packard Fellowship in Science and
Engineering. L.J. acknowledges support from the
Alfred P. Sloan Foundation, the Packard
Foundation, the Multidisciplinary University
Research Initiative (MURI), and the DARPA Quiness
program. C.L.Z. acknowledges National Basic
Research Program of China (Grant Nos.
2011CB921200 and 2011CBA00200). F.M. acknowledges
ERC OPTOMECH and DARPA ORCHID. The authors thank
Dr. Michel H. Devoret and Dr. Michael Hatridge
for providing a prototype 3D microwave cavity.

\vbox{}

\noindent \textbf{Author contributions}\\X.Z.
prepared the samples, performed the measurements
and data analysis. C.L.Z. and L.J. conceived the
idea. C.L.Z., L. J. and F.M. provided theoretical
analysis. N.Z. assisted with the measurements.
L.J. and H.X.T. supervised the project. X.Z.,
C.L.Z., F.M., L.J., and H.X.T. wrote the
manuscript.

\vbox{}

\noindent \textbf{Additional information}\\
 Supplementary Information is available in the online version of the
paper. Reprints and permissions information is available online at
www.nature.com/reprints.Correspondence and requests for materials
should be addressed to H.X.T. (hong.tang@yale.edu).

\vbox{}

\noindent \textbf{Competing financial interests}\\
 The authors declare no competing financial interests.

\clearpage{}\newpage{}

\newpage{}

\onecolumngrid

\global\long\def\thefigure{S\arabic{figure}}\setcounter{figure}{0}
 \global\long\def\thepage{S\arabic{page}} \setcounter{page}{0}
 \global\long\def\theequation{S\arabic{equation}} \setcounter{equation}{0}
\global\long\def\thefootnote{S\arabic{footnote}}\setcounter{footnote}{0}

\begin{center}
\textbf{\textsc{\LARGE{}{}{}Supplementary Information}}{\LARGE{}{}
}
\par\end{center}

\tableofcontents{}

\newpage{}

\section{Microwave photon-magnon strong coupling}

The interaction between microwave photon and
magnon (single mode) can be described by the
Hamiltonian with the rotating-wave approximation
\begin{equation}
\mathcal{\hat{H}}/\hbar=\omega_{a}\hat{a}^{\dag}\hat{a}+\omega_{m}\hat{m}^{\dag}\hat{m}+g(\hat{a}^{\dag}\hat{m}+\hat{a}\hat{m}^{\dag}),\label{eq:HamiltonianS}
\end{equation}

\noindent where $\hat{a}^{\dag}$ ($\hat{a}$) is
the creation (annihilation) operator for the
microwave photon at frequency $\omega_{a}$.For
the uniform magnon mode, the collective spin
excitations are approximately represented by the
Boson operator $\hat{m}^{\dag}$($\hat{m}$) with
Holstein--Primakoff approximation \footnotemark
and $\omega_{m}$ is the magnon frequency.
\footnotetext{Holstein, T. and Primakoff, H.
\textit{Phys. Rev.} \textbf{58}, 1098 (1940).}
The coupling strength $g$ between the two systems
is \footnotemark:
\begin{equation}
g=\frac{\eta}{2}\gamma\sqrt{\frac{\hbar\omega\mu_{0}}{V_{a}}}\sqrt{2Ns},\label{eq:g}
\end{equation}

\footnotetext{Zhang, X., Zou, C.-L., Jiang, L.,
and Tang, H. X. \textit{Phys. Rev. Lett.}
\textbf{113}, 156401 (2014).}

\noindent where $\omega$ is the resonance frequency, $\gamma$ is
the gyromagnetic ratio, $V_{a}$ is the modal volume of the microwave
cavity resonance, $\mu_{0}$ is the vacuum permeability, $N$ is the
total number of spins, and $s=\frac{5}{2}$ is the spin number of
the ground state Fe$^{3+}$ ion in YIG. The coefficient $\eta\leq1$
describes the spatial overlap and polarization matching conditions
between the microwave field and the magnon mode, which can be explicitly
written as:
\begin{equation}
\eta^{2}=\frac{(\overrightarrow{h}(\mathbf{r})\cdot\overrightarrow{e_{x}})^{2}+(\overrightarrow{h}(\mathbf{r})\cdot\overrightarrow{e_{y}})^{2}}{\mathrm{max}\{|\overrightarrow{h}(\mathbf{r})|^{2}\}},\label{eq:eta}
\end{equation}
where
$\mathrm{max}\{|\overrightarrow{h}(\mathbf{r})|^{2}\}$
is the maximum magnetic field intensity of the
cavity mode, and $\overrightarrow{h}(\mathbf{r})$
is the magnetic field amplitude at the location
($\mathbf{r}$) of the YIG sphere,
$\overrightarrow{e_{j}}$ with $j=x,y,z$ are unit
vectors and $\overrightarrow{e_{z}}$ is along the
bias field direction.

\section{Reflection spectrum}

\label{sec:ReflectionSpectrum}In our experiment,
we measure the reflection spectrum under
continues driving and the emission after pulsed
excitations of the microwave cavity, which
correspond to the eigenmode spectrum and the
transient cavity ouput, respectively. In general,
the equation of motion for the cavity photon is
\begin{equation}
\frac{d}{dt}\hat{a}=-i[\hat{a},\mathcal{\hat{H}}]-i\sqrt{2\kappa_{a,1}}E_\mathrm{in}(t)e^{-i\omega_{l}t},\label{eq:EqMotionCavityInputOutput}
\end{equation}
where $\mathcal{\hat{H}}$ is the system
Hamiltonian, $\kappa_{a,1}$ is the cavity
external coupling rate to the coaxial probe,
$E_{in}(t)$ and $\omega_{l}$ are the amplitude
and frequency of the input microwave.

In the rotating frame of $\omega_{l}$, the microwave electric field
collected by the detector reads
\begin{equation}
E_\mathrm{out}(t)=-E_\mathrm{in}(t)+i\sqrt{2\kappa_{a,1}}\hat{a}(t)\label{eq:input-output}
\end{equation}

For a continuous stable input $E_{in}(t)=E_{in}$,
we measure the stationary spectra with the
normalized spectrum expressed as
\begin{equation}
r(\omega_{l})=-1+i\sqrt{2\kappa_{a,1}}\frac{\hat{a}(\omega_{l})}{E_\mathrm{in}}.\label{eq:reflection}
\end{equation}

For the memory operation, the retrieval of the
microwave pulse is measured after the input pulse
is stored, and the detected intensity is
\begin{equation}
I(t)=2\kappa_{a,1}|\left<\hat{a}(t)\right>|^{2}.\label{eq:emission}
\end{equation}

The efficiency of the MGM, defined as the energy
ratio of the retrieval pulse to the input pulse,
can be expressed as
\begin{equation}
\zeta=\frac{\int I(t)dt}{\int
 I_\mathrm{in}(t)dt}.\label{eq:efficiency}
\end{equation}

\section{Dark magnon modes}

In general, the Hamiltonian of multiple linearly
coupled magnon modes and the cavity mode reads:
\begin{equation}
\mathcal{\hat{H}}/\hbar=\omega_{a}\hat{a}^{\dagger}\hat{a}+\sum_{j=1}^{N}(\omega_{j}\hat{m}_{j}^{\dagger}\hat{m}_{j}+g_{j}\hat{a}^{\dagger}\hat{m}_{j}+g_{j}^{*}\hat{a}\hat{m}_{j}^{\dagger}).
\end{equation}
When deriving the equations of motion, both the
magnon dissipation and the cavity dissipation
(including the external coupling) are considered.
In principle, the input-output theory shows that
this gives a damping term as well as a
fluctuation (including vacuum and thermal) term.
To keep the notation simple, we will, from now
on, use symbols like $a$ and $m_{j}$ to
\emph{refer to the expectation values of the
underlying operators}. Therefore, the fluctuation
terms will be dropped. This is sufficient to
obtain the linear scattering behavior of the
system and to discuss its eigenmodes. With this
proviso, the equations of motion for the cavity
mode and the magnon modes are:
\begin{eqnarray}
\frac{d}{dt}a & = & (-i\omega_{a}-\kappa_{a})a-i\sum_{j}g_{j}m_{j},\label{eq:EqMotionCavity}\\
\frac{d}{dt}m_{j} & = & (-i\omega_{j}-\kappa_{j})m_{j}-ig_{j}^{*}a.\label{eq:EqMotionMagnon}
\end{eqnarray}
Here, $\kappa_{a}$ and $\kappa_{j}$ are the total
decay rates of the cavity photon and the magnon
modes, respectively, with $\kappa_{a}$ including
both the internal dissipation ($\kappa_{a,0}$) as
well as the coupling to the coaxial probe
($\kappa_{a,1}$).

\subsection{Two Magnon Modes}

As described in the main text, we first study the
case with two magnon modes. For simplicity, let
the two magnon modes be on-resonance with the
cavity mode ($\omega_{1}=\omega_{2}=\omega_{m}$).
Then the dynamics of the cavity field is
\begin{eqnarray}
\frac{d}{dt}a & = & (-i\omega_{a}-\kappa_{a})a-i(g_{1}m_{1}+g_{2}m_{2}).
\end{eqnarray}
The cavity mode couples to the collective mode $B=(g_{1}m_{1}+g_{2}m_{2})/\sqrt{\left|g_{1}\right|^{2}+\left|g_{2}\right|^{2}}$
as
\begin{eqnarray}
\frac{d}{dt}B & = & (-i\omega_{m}-\kappa_{m})B-i\sqrt{\left|g_{1}\right|^{2}+\left|g_{2}\right|^{2}}a.
\end{eqnarray}
And the orthogonal mode $D=(g_{2}^{*}m_{1}-g_{1}^{*}m_{2})/\sqrt{\left|g_{1}\right|^{2}+\left|g_{2}\right|^{2}}$
satisfies
\begin{eqnarray}
\frac{d}{dt}D & = & (-i\omega_{m}-\kappa_{m})D-i(g_{1}^{*}g_{2}^{*}-g_{2}^{*}g_{1}^{*})a/\sqrt{\left|g_{1}\right|^{2}+\left|g_{2}\right|^{2}}\\
 & = & (-i\omega_{m}-\kappa_{m})D.
\end{eqnarray}
Therefore, $D$ is completely isolated from the
cavity mode, representing the dark mode that
cannot be detected from the reflection spectrum
of the microwave cavity.

\subsection{Generalization to Multiple Magnon Modes}

For multiple magnon modes $N\geq2$ and
$\omega_{j}=\omega_{1}$ for $j=2,\ldots,\ N$, the
normalized bright mode is
\begin{equation}
B=\sum_{j=1}^{N}g_{j}m_{j}/\sqrt{\sum_{j=1}^{N}\left|g_{j}\right|^{2}}.
\end{equation}
Then we have
\begin{eqnarray}
\frac{d}{dt}a & = & (-i\omega_{a}-\kappa_{a})a-i\sqrt{\sum_{j=1}^{N}\left|g_{j}\right|^{2}}B.
\end{eqnarray}
If the magnons are identical, the bright mode is a collective mode
of all the magnon modes with the coupling strength being enhanced
by a factor of
\begin{equation}
f=\sqrt{\sum_{j}\left|g_{j}\right|^{2}}/g_{1}=\sqrt{N}.
\end{equation}
There are $N-1$ other modes, orthogonal to the
bright mode and decoupled from the cavity.

\subsection{Temporal Dark Magnon Mode}

In the above analysis, all the magnon modes are
on resonance with the cavity mode, and therefore
the bright and dark modes are all the eigenmodes
of the Hamiltonian. When the magnon modes are
tuned off but close to resonance, situation would
be different. As an example, we consider the case
of two magnon modes. If the two magnon modes are
detuned from the cavity mode as
\begin{align}
\omega_{1} & =\omega_{a}+\Delta\omega/2,\\
\omega_{2} & =\omega_{a}-\Delta\omega/2,
\end{align}

\noindent then we can define the temporal bright
mode at time $t$ as (assuming
$g_{1}=g_{2}=g_{m}$)
\begin{equation}
\widetilde{B}(t)=\frac{1}{\sqrt{2}}e^{-i\omega_{a}t}\left(m_{1}e^{-i\Delta\omega
t/2}+m_{2}e^{i\Delta\omega t/2}\right).
\end{equation}

\noindent At $t=0$, this mode is
$B=\frac{1}{\sqrt(2)}(m_1+m_2)$, which is the
same as static bright mode discussed above. But
at $t=\frac{\pi}{\Delta\omega}$, the mode evolves
to
\begin{equation}
\widetilde{B}(\frac{\pi}{\Delta\omega})=\frac{-i}{\sqrt{2}}e^{-i\omega_{a}\pi/\Delta\omega}\left(m_{1}-m_{2}\right)=-ie^{-i\omega_{a}\pi/\Delta\omega}D,
\end{equation}
\noindent which is the dark mode that decouples
from the cavity. Therefore, in this scenario, the
temporal bright and dark magnon modes are not
eigenmodes of the system and they are
inter-convertible through time evolution.

\section{Magnon gradient memory}

\subsection{Intuitive Explanation}

To realize a magnon memory device using the
temporal magnon dark mode, we want: (1) the input
photon can be converted to the collective magnon
bright mode as quickly as possible, and therefore
the photon energy will not dissipate too much due
to the cavity intrinsic loss; (2) the bright mode
can convert to the dark mode once the input
photon is converted to the bright mode; (3) the
retrieval of the photon is predictable and
pre-programmable.

To fulfill these requirements, the system should
consist of multiple magnon modes with equal
detuning. Assuming the lifetime of the magnon and
the cavity photon are $\tau_{m}=1/2\kappa_{m}$
and $\tau_{a}=1/2\kappa_{a}$, respectively, the
retrieval period $T=\pi/\Delta\omega$, the
conversion time from the bright mode to the dark
modes $t_{B\rightarrow D}=T/N$ with $N$ being the
total number of the magnon modes, and the
conversion time from the cavity photon to the
magnon bright mode $t_{a\rightarrow
B}=\pi/\sqrt{N}g$, then we will have the
following constraints
\begin{align}
T & \leq\tau_{m},\\
t_{B\rightarrow D} & \geq t_{a\rightarrow B},\\
t_{a\rightarrow B} & \ll\tau_{a},
\end{align}
which can be rewritten as
\begin{align}
\Delta\omega & \geq\frac{\pi}{\tau_{m}}=2\pi\times\kappa_{m},\label{eq:magnonloss_small}\\
g & \geq\sqrt{N}\Delta\omega\geq2\pi\sqrt{N}\times\kappa_{m},\\
g & \gg\frac{2\pi}{\sqrt{N}}\times\kappa_{a,0}.\label{eq:g_muchlarger}
\end{align}
A highly efficient magnon gradient memory requires the cooperativity
$C=\frac{g^{2}}{\kappa_{a,0}\kappa_{m}}\gg4\pi^{2}$. In our experiment,
the hybrid YIG sphere--3D cavity structure can have a cooperativity
$C$ as large as $10^{4}$, and therefore it is very promising for
memory applications.

\subsection{Dynamics}

In an ideal magnon gradient memory, we have
\begin{equation}
\omega_{j}=\omega_{a}-\frac{N-1}{2}\Delta\omega+j\Delta\omega
\end{equation}
for the uniform magnon mode in the $j-$th YIG sphere. Since
\begin{equation}
\frac{d}{dt}m_{j}=(-i\omega_{j}-\kappa_{j})m_{j}-ig_{j}^{*}a,
\end{equation}
we have the formal solution for the magnon time-evolution (where $m_{j}(0)=0$):
\begin{equation}
m_{j}(t)=-ig_{j}^{*}\int_{0}^{t}a(\tau)e^{(-i\omega_{j}-\kappa_{j})(t-\tau)}d\tau.
\end{equation}
Then, the cavity dynamics is
\begin{eqnarray}
\frac{d}{dt}a & = & (-i\omega_{a}-\kappa_{a})a-\sum_{j}\left|g_{j}\right|^{2}\int_{0}^{t}a(\tau)e^{(-i\omega_{j}-\kappa_{j})(t-\tau)}d\tau\\
 & = & (-i\omega_{a}-\kappa_{a})a-\int_{0}^{t}a(\tau)\sum_{j}\left|g_{j}\right|^{2}e^{(-i\omega_{j}-\kappa_{j})(t-\tau)}d\tau.
\end{eqnarray}
For $N$ identical YIG spheres with $g_{j}=g$ and $\kappa_{j}=\kappa_{m}$,
we have
\begin{eqnarray}
\sum_{j}\left|g_{j}\right|^{2}e^{(-i\omega_{j}-\kappa_{j})(t-\tau)} & = & \sum_{j}\left|g_{j}\right|^{2}e^{(-i\omega_{a}-i\frac{N-1}{2}\Delta\omega-ij\Delta\omega-\kappa_{j})(t-\tau)}\nonumber \\
 & = & \left|g\right|^{2}e^{(-i\omega_{a}-i\frac{N-1}{2}\Delta\omega-\kappa_{m})(t-\tau)}\sum_{j=1}^{N}e^{-ij\Delta\omega(t-\tau)}\nonumber \\
 & = & \left|g\right|^{2}e^{(-i\omega_{a}-\kappa_{m})(t-\tau)}\frac{\mathrm{sin}[\frac{N\Delta\omega(t-\tau)}{2}]}{\mathrm{sin}[\frac{\Delta\omega(t-\tau)}{2}]}.
\end{eqnarray}
Finally, the cavity dynamics is determined by the following equation
\begin{eqnarray}
\frac{d}{dt}a(t) & = &
(-i\omega_{a}-\kappa_{a})a(t)-\left|g\right|^{2}\int_{0}^{t}a(\tau)e^{(-i\omega_{a}-\kappa_{m})(t-\tau)}\frac{\mathrm{sin}[\frac{N\Delta\omega(t-\tau)}{2}]}{\mathrm{sin}[\frac{\Delta\omega(t-\tau)}{2}]}d\tau.\label{eq:dynamics}
\end{eqnarray}

\subsection{Asymptotic Solution}

\begin{figure}
\begin{centering}
\includegraphics[width=10cm]{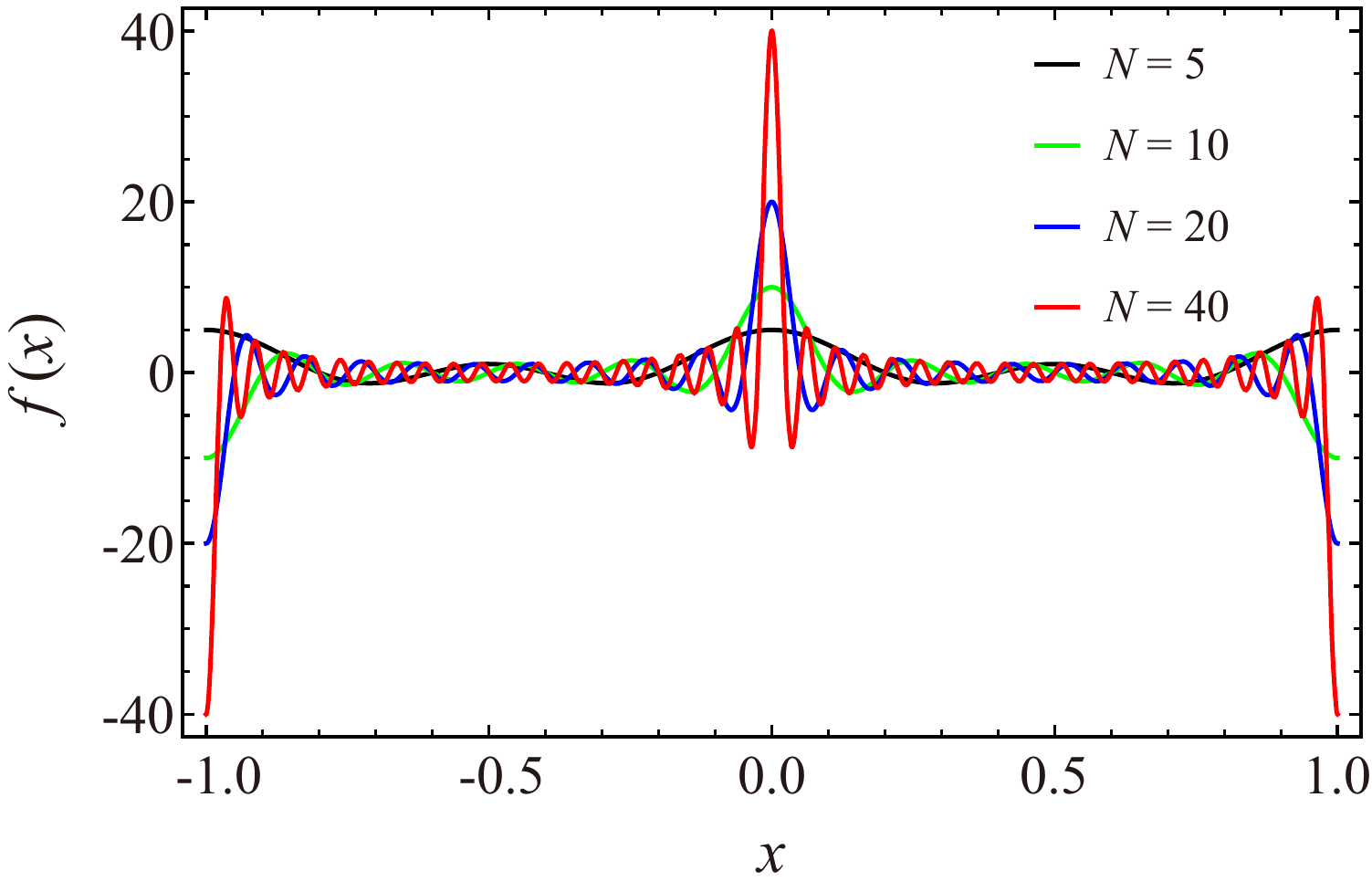}
\par\end{centering}
\protect\protect\protect\caption{The plot of the
function
$f(x)=\frac{\mathrm{sin}[\frac{N}{2}x]}{\mathrm{sin}[\frac{1}{2}x]}.$}
\label{fig:S1}
\end{figure}

As shown by Fig.\,\ref{fig:S1}, we can take the
approximation
\begin{equation}
\frac{\mathrm{sin}[\frac{N\Delta\omega(t-\tau)}{2}]}{\mathrm{sin}[\frac{\Delta\omega(t-\tau)}{2}]}\approx2\pi\sum_{l}(-1)^{l(N-1)}\delta[\Delta\omega(t-\tau-l\frac{2\pi}{\Delta\omega})]
\end{equation}
for $N\gg1$, where $\delta(x)$ is the Dirac delta function.

Therefore, we can describe the cavity dynamics
for $N\gg1$ as
\begin{eqnarray}
\frac{d}{dt}a(t) & = & (-i\omega_{a}-\kappa_{a})a(t)-2\pi\left|g\right|^{2}\int_{0}^{t}a(\tau)e^{(-i\omega_{a}-\kappa_{m})(t-\tau)}\sum_{l=-\infty}^{\infty}(-1)^{l(N-1)}\delta[\Delta\omega(t-\tau-l\frac{2\pi}{\Delta\omega})]d\tau\nonumber \\
 & = & (-i\omega_{a}-\kappa_{a})a(t)-\frac{\pi\left|g\right|^{2}}{\Delta\omega}a(t)-\sum_{l\neq0}(-1)^{l(N-1)}\frac{2\pi\left|g\right|^{2}}{\Delta\omega}a(t-l\frac{2\pi}{\Delta\omega})e^{(-i\omega_{a}-\kappa_{m})(t-\tau)}.
\end{eqnarray}

For $t<\frac{2\pi}{\Delta\omega}$, the cavity photon experiences
single exponential decay
\begin{equation}
\frac{d}{dt}a(t)=(-i\omega_{a}-\kappa_{a})a(t)-\frac{\pi\left|g\right|^{2}}{\Delta\omega}a(t),
\end{equation}
with solution
\begin{equation}
a(t)=a(0)e^{-(i\omega_{a}+\kappa_{a}+\frac{\pi\left|g\right|^{2}}{\Delta\omega})t}.\label{eq:input}
\end{equation}
We can see that the cavity photon decay is enhanced by $\frac{\pi\left|g\right|^{2}}{\Delta\omega}$.

For $\frac{2\pi}{\Delta\omega}\leq t\leq2\frac{2\pi}{\Delta\omega}$,
we have
\begin{eqnarray}
\frac{d}{dt}a(t) & = & (-i\omega_{a}-\kappa_{a}-\frac{\pi\left|g\right|^{2}}{\Delta\omega})a(t)+\frac{2\pi\left|g\right|^{2}}{\Delta\omega}a(t-\frac{2\pi}{\Delta\omega})e^{(-i\omega_{a}-\kappa_{m})\frac{2\pi}{\Delta\omega}},\\
 & = & (-i\omega_{a}-\kappa_{a}-\frac{\pi\left|g\right|^{2}}{\Delta\omega})a(t)+\frac{2\pi\left|g\right|^{2}}{\Delta\omega}a(0)e^{(-i\omega_{a}-\kappa_{m})\frac{2\pi}{\Delta\omega}}e^{-(i\omega_{a}+\kappa_{a}+\frac{\pi\left|g\right|^{2}}{\Delta\omega})(t-\frac{2\pi}{\Delta\omega})}.
\end{eqnarray}
Therefore, the cavity photons show a revival due to the energy of
magnons coupling back to the cavity
\begin{equation}
a(t)=-\frac{2\pi\left|g\right|^{2}}{\Delta\omega}a(0)e^{(-i\omega_{a}-\kappa_{m})\frac{2\pi}{\Delta\omega}}(t-\frac{2\pi}{\Delta\omega})e^{-(i\omega_{a}+\kappa_{a}+\frac{\pi\left|g\right|^{2}}{\Delta\omega})(t-\frac{2\pi}{\Delta\omega})}.\label{eq:asymptotic}
\end{equation}

\subsection{Scattering Picture}

Above, we have adopted a direct solution of the
temporal equations of motion. Alternatively, one
can also discuss the MGM as a scattering problem
in frequency space. We employ the equations of
motion, Eqs.\,(\ref{eq:EqMotionCavity}) and
(\ref{eq:EqMotionMagnon}), supplement them with
the input field (as in
Eq.\,(\ref{eq:EqMotionCavityInputOutput})), and
obtain, in frequency space:

\begin{align}
-i\omega a & =(-i\omega_{a}-\kappa_{a})a-i\sum_{j}g_{j}m_{j}-i\sqrt{2\kappa_{a,1}}E_{{\rm in}}\\
-i\omega m_{j} & =(-i\omega_{j}-\kappa_{m})m_{j}-ig_{j}^{*}a\,.
\end{align}
Here the functions $a$, $m_{j}$, and $E_{{\rm in}}$ are understood
to be functions of the frequency $\omega$.

These equations can be solved by eliminating
$m_{j}$ in favor of $a$ and then applying the
input-output relation
Eq.\,(\ref{eq:input-output}). This leads to the
following result for the frequency-dependent
reflection amplitude that gives $E_{{\rm
out}}(\omega)=r(\omega)E_{{\rm in}}(\omega)$:

\begin{equation}
r(\omega)=\frac{\left[\omega_{a}-\omega+\Sigma(\omega)\right]+i(\kappa_{a,1}-\kappa_{a,0})}{-\left[\omega_{a}-\omega+\Sigma(\omega)\right]+i(\kappa_{a,1}+\kappa_{a,0})}\,.
\end{equation}
At this point we have introduced the ``MGM self-energy'' that describes
the collective effects of all the magnon modes acting on the cavity
mode:

\begin{equation}
\Sigma(\omega)=\sum_{j}\frac{\left|g_{j}\right|^{2}}{\omega-\omega_{j}+i\kappa_{m}}\,.
\end{equation}
The real part of $\Sigma$ describes the effective frequency shift,
while the imaginary part describes additional damping (induced by
the magnon mode damping). The reflection amplitude can be decomposed
into magnitude and phase shift:

\begin{equation}
r(\omega)=|r(\omega)|e^{i\theta(\omega)}\,.
\end{equation}
In the ideal case without intrinsic losses ($\kappa_{a,0}=\kappa_{m}=0)$,
we have $\left|r\right|=1$. The time-delay of a scattered wave-packet
is determined by the derivative of the phase shift with respect to
frequency:

\begin{equation}
\tau(\omega)=\frac{\partial\theta(\omega)}{\partial\omega}\,.
\end{equation}
The most important necessary condition for a
useful memory is that this time-delay be constant
over the bandwidth interval $N\Delta\omega$, i.e.
the slope of $\theta$ should be constant. In the
limit of small magnon-cavity coupling
$g\rightarrow0$ (and no intrinsic losses), each
magnon resonance leads to a step of $2\pi$ in the
phase shift $\theta$. A finite $g$ rounds off
these steps. In the vicinity of each resonance,
we can then approximate $\Sigma\approx
g^{2}/(\omega-\omega_{j})$, which leads to a
phase shift rising like
$\theta(\omega)-\theta(\omega_{j})=2\kappa_{a,1}(\omega-\omega_{j})/g^{2}$.
In order to have a constant overall slope of
$\theta$, we have to match this to the slope
$T=2\pi/\Delta\omega$ that is dictated by the
spacing of resonances and which corresponds to
the ideal storage time. That leads to the
critical coupling condition (without intrinsic
losses)

\begin{equation}
\frac{\pi g^{2}}{\Delta\omega}=\kappa_{a,1}\,.
\end{equation}
We note that this condition remains true (in the
present form) if intrinsic cavity losses are also
incorporated, i.e. when $\kappa_{a,0}\neq0$. The
contribution from $\kappa_{a,0}$ cancels when
deriving the condition. We note, however, that
for finite $\kappa_{a,0}$ this condition slightly
differs from the critical coupling condition
derived below from demanding zero reflection of
the input pulse. This is because, in general, for
finite $\kappa_{a,0}$, the form of the critical
coupling condition depends on the precise
physical condition that is imposed. Still, one
needs ideally $\kappa_{a,0}\ll\kappa_{a,1}$ to
suppress unwanted losses and to avoid a resonance
structure showing up in the magnitude
$\left|r\right|$ of the reflection. In addition,
the cavity mode acts like a filter, which should
be broad enough to cover the whole magnon
spectrum, i.e.

\begin{equation}
\kappa_{a,1}\gg N\Delta\omega\,.
\end{equation}
Taking these two conditions together also implies
$g\gg\kappa_{a,1}/\sqrt{N}$, as stated already in
Eq.\,(\ref{eq:g_muchlarger}). In addition, to
keep the magnon losses small (ensuring
$\left|r\right|$ close to $1$), one needs
$\kappa_{m}\ll\Delta\omega/2\pi$, as stated in
Eq.\,(\ref{eq:magnonloss_small}).

\begin{figure}[htpb]\begin{centering}
\includegraphics[width=16cm]{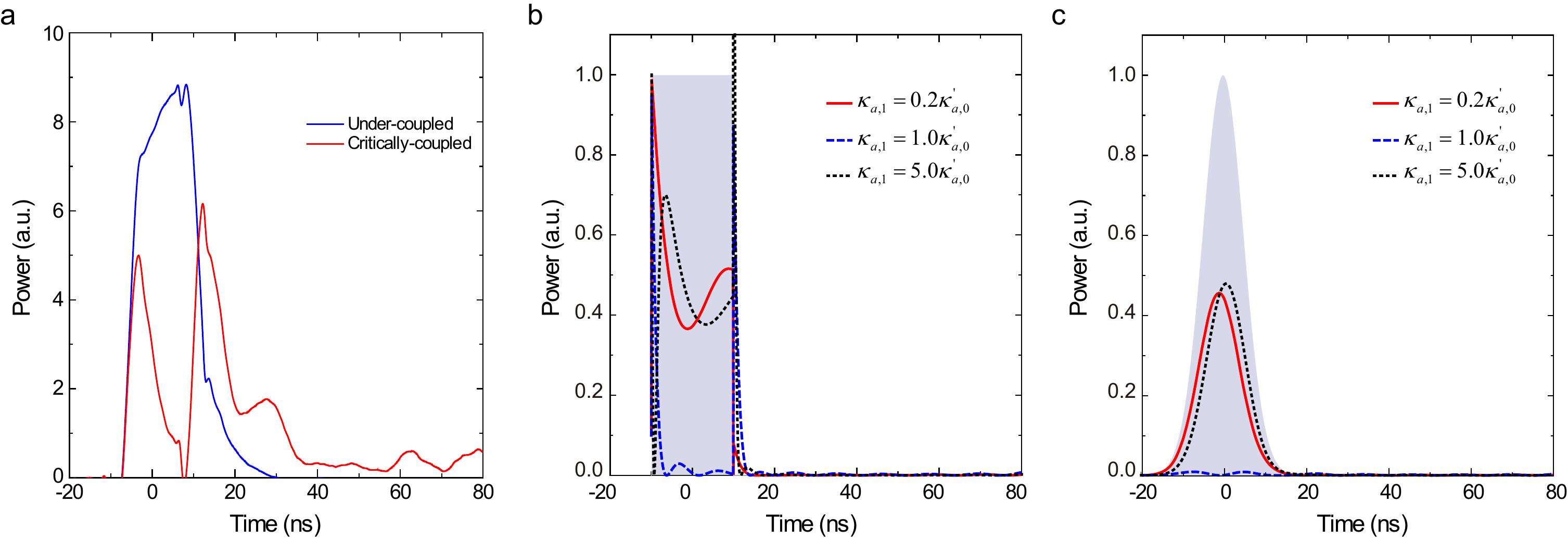}\par\end{centering}
\protect\caption{Input reflection at various
coupling conditions. a, Measured input
reflections for microwave pulses to the microwave
cavity with under-coupled
($\kappa_{a,1}<\kappa'_{a,0}$, blue curve) and
critically coupled
($\kappa_{a,1}\approx\kappa'_{a,0}$, red curve)
probes, respectively. b, The calculated cavity
reflection for a rectangular pulse input (shadow
region) at various coupling conditions. c, The
calculated cavity reflection for a Gaussian input
pulse using the same parameters as in b.}
\label{fig:S2}
\end{figure}

\subsection{Critical coupling condition}

From Eq.\,(\ref{eq:input}), the microwave pulse
input into the cavity couples to the external
coupling channel (rate $\kappa_{a,1}$) and the
temporal bright mode
($\frac{\pi\left|g_{0}\right|^{2}}{\Delta\omega}$),
and also dissipates ($\kappa_{a,0}$) due to
intrinsic radiation and absorption losses.
Therefore, the full internal loss of the cavity
is actually
$\kappa'_{a,0}=\kappa_{a,0}+\frac{\pi\left|g_{0}\right|^{2}}{\Delta\omega}$.
To suppress the reflection of the input pulse,
the critical coupling (impedance matching)
condition
$\kappa_{a,1}=\kappa'_{a,0}=\kappa_{a,0}+\frac{\pi\left|g_{0}\right|^{2}}{\Delta\omega}$
is desired. The storage and retrieval efficiency
of the MGM can be greatly affected by the
critical coupling condition of the MGM with the
coaxial probe. If the MGM is under-coupled, a
large portion of the energy will be reflected as
the signal gets in and out of the cavity, and as
a result the MGM efficiency becomes low.

In our measurements, we did observe a strong
reflection peak at the time of input for the
undercoupling situation, and such reflection peak
can be eliminated after adjusting the external
coupling rate to meet the critical coupling
condition, as plotted in Fig.\,\ref{fig:S2}a.
Such an observation agrees well with the
theoretical predication (Fig.\,\ref{fig:S2}b).
Note that there is still some residual reflection
for the critical coupling condition, which is
attributed to the step response of the MGM to the
rectangular input pulse (both the rising and
falling edges). This imperfection can be
suppressed by choosing Gaussian pulse inputs
(Fig.\,\ref{fig:S2}c).

\subsection{Photon Retrieval}

\begin{figure}[htpb]
\begin{centering}
\includegraphics[width=16cm]{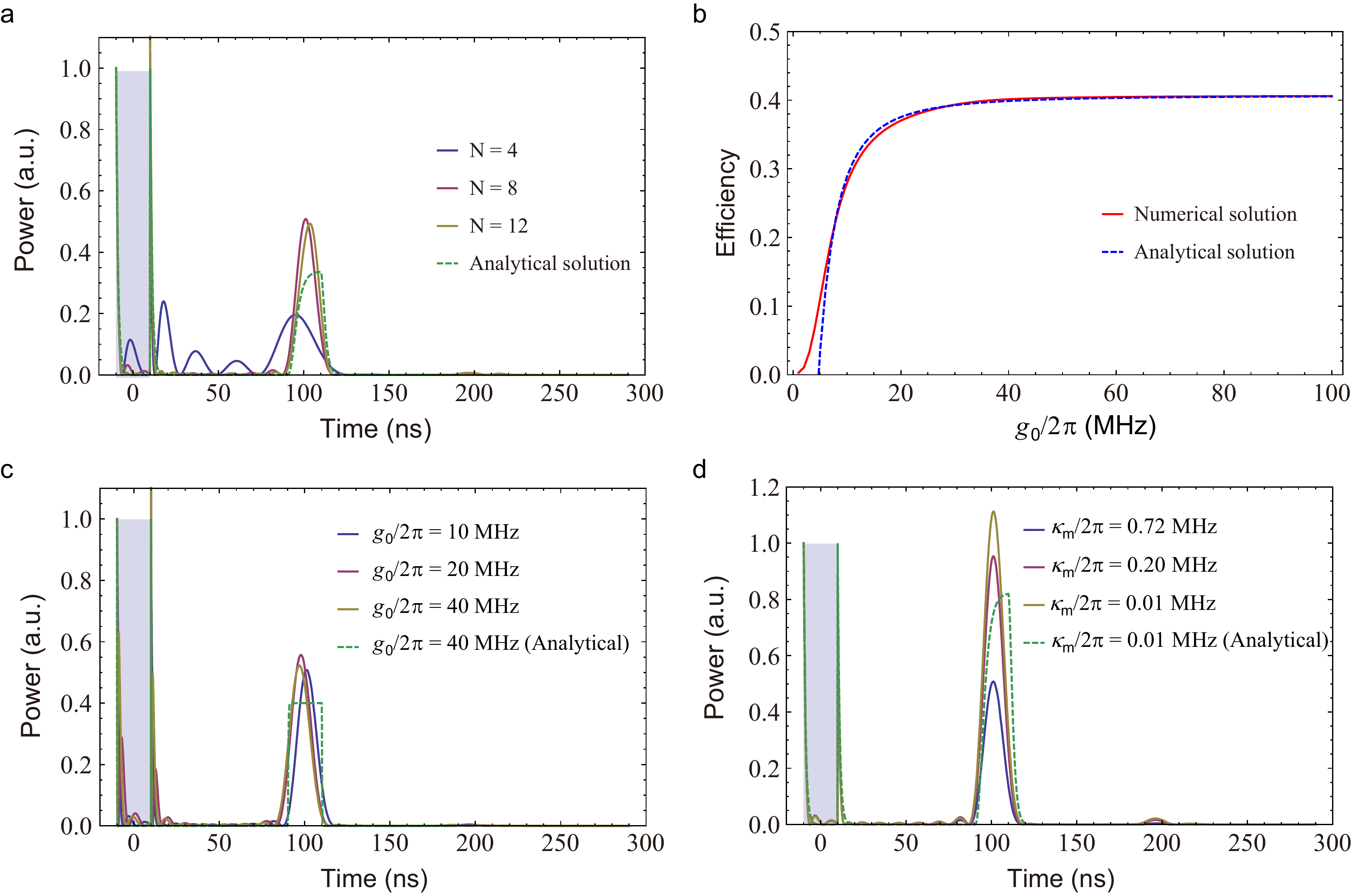}
\end{centering}
\caption{a, Calculated dynamics of the MGMs with
different YIG sphere numbers ($N$). b, Comparison
of the numerical and analytical solutions at
various coupling strengths. c \& d, dynamics of
the MGM with various coupling strengths and
magnon dissipation rates, respectively. For each
plot, the solid and dashed curves represent the
numerical and analytical solutions, respectively.
All parameters (beside the varying parameters)
are taken from the experiments: $N_{s}=8$,
$g_0=2\pi\times10\ \mathrm{MHz}$,
$\Delta\omega=2\pi\times10\ \mathrm{MHz}$,
$\kappa_{a,0}=2\pi\times3\ \mathrm{MHz}$,
$\kappa_{m}=2\pi\times0.72\ \mathrm{MHz}$,
$\kappa_{a,1}=\kappa_{a,0}+\frac{\pi\left|g_{0}\right|^{2}}{\Delta\omega}$
(critical coupling), and the input pulse duration
is $t_{p}=20$ ns.} \label{fig:S3}

\vskip 10mm

\begin{centering}
\includegraphics[width=16cm]{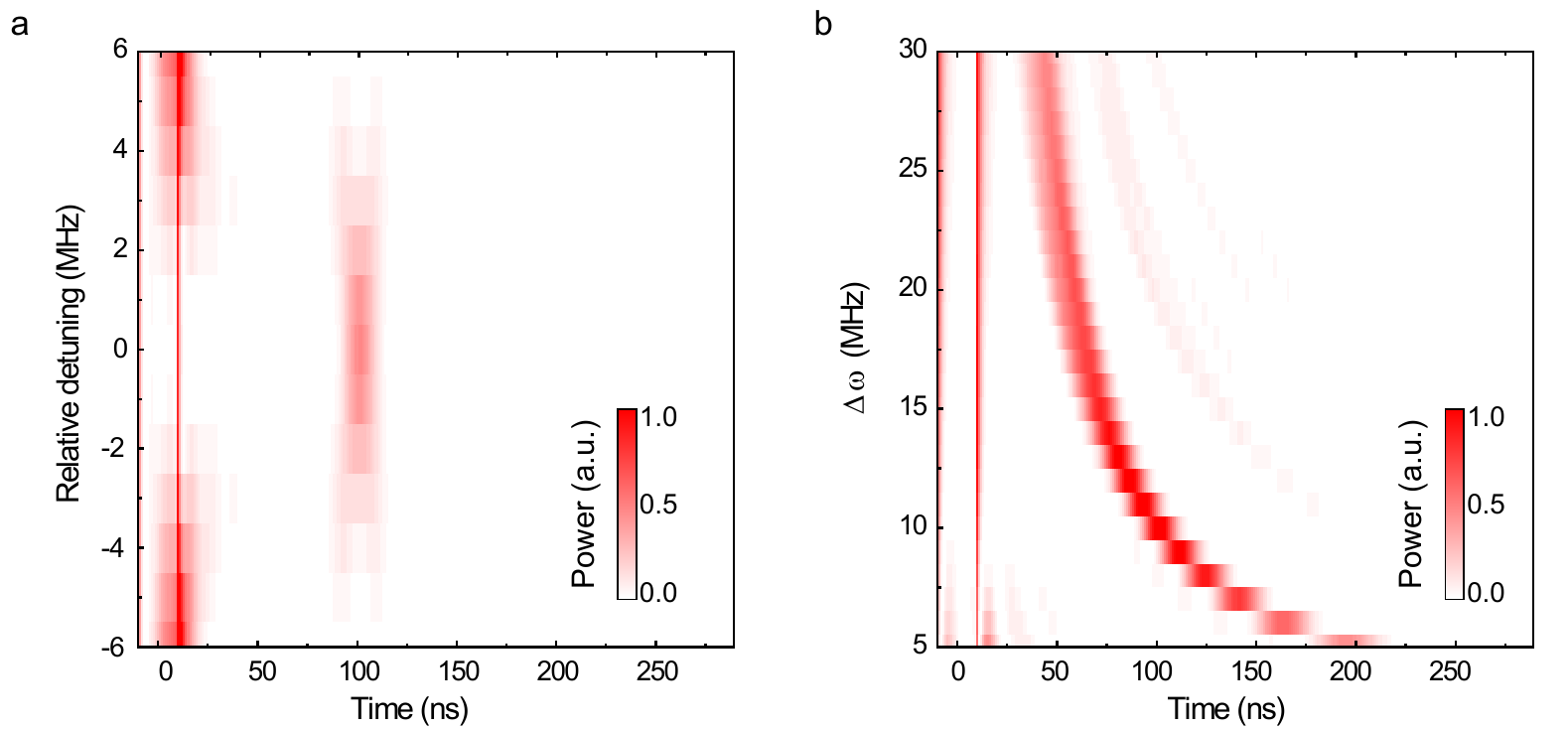}
\end{centering}
\caption{The calculated reflection signal
dynamics at (a): various input microwave
detunings
$(\omega_{\mathrm{in}}-\omega_{a})/\Delta\omega$,
and (b): various magnon frequency gradients
$\Delta\omega$. In the calculation, we assume all
the magnon modes are identical, have uniform
coupling strength, and are evenly distributed.
Other parameters are from our experiment:
$N_{s}=8$, $g_0=2\pi\times10\ \mathrm{MHz}$,
$\Delta\omega=2\pi\times10\ \mathrm{MHz}$,
$\kappa_{a,0}=2\pi\times3\ \mathrm{MHz}$,
$\kappa_{m}=2\pi\times0.72\ \mathrm{MHz}$,
$\kappa_{a,1}=\kappa_{a,0}+\frac{\pi\left|g_{0}\right|^{2}}{\Delta\omega}$
(critical coupling) and the input pulse duration
is $t_{p}=20$ ns.}\label{fig:S4}
\end{figure}

From Eq.\,(\ref{eq:asymptotic}), after one period
$T=\frac{2\pi}{\Delta\omega}$ the stored energy
will couple back to the cavity, which leads to
the retrieval of the photons. We can solve for
the detected microwave amplitude by the
input-output formula
Eq.\,(\ref{eq:input-output}), with the input
microwave also taken into account. Substituting
the experiment parameters into the equation of
the cavity field, with the input rectangular
pulse duration $t_{p}=20$ ns, we obtained the
dynamics in Fig.\,\ref{fig:S3}a. The analytical
and numerical results agree well with each other.
With the given parameters, the numerical solution
shows a memory efficiency of $\zeta=0.33$, which
is the integrated total output energy in the
first retrieval peak as compared with that of the
input pulse.

From the asymptotic solution, we can derive the
efficiency of the magnon gradient memory as
\begin{equation}
\zeta\approx e^{-2\kappa_{m}\frac{2\pi}{\Delta\omega}}\left[1-\frac{1}{t_{p}\left(\kappa_{a,0}+\frac{\pi\left|g\right|^{2}}{\Delta\omega}\right)}\right]\left(\frac{\frac{\pi\left|g\right|^{2}}{\Delta\omega}}{\kappa_{a,0}+\frac{\pi\left|g\right|^{2}}{\Delta\omega}}\right)^{2}
\end{equation}
for a rectangular input pulse under critical
coupling condition (which demands zero
reflection). Note that the pulse duration is
limited by the external coupling rate:
$t_p>1/\kappa_{a,1}=1/(\kappa_{a,0}+\frac{\pi|g|^2}{\Delta\omega})$.

In Fig.\,\ref{fig:S3}b, we plot the efficiency as
a function of the coupling strength $g_{0}$. A
comparison of the numerical results with the
asymptotic solutions shows that they agree well
with each other for $g/2\pi\geq7$ MHz. The
efficiency saturates to
$\zeta=e^{-2\kappa_{m}\frac{2\pi}{\Delta\omega}}\approx0.403$
for
$\frac{\pi\left|g_{0}\right|^{2}}{\Delta\omega}\gg\kappa_{a,0}$,
which is limited by the intrinsic loss of the
magnon. Figure \ref{fig:S3}c plots the behavior
of the memory for various coupling strengths,
which shows the saturated retrieval peak at 100
ns. If we can reduce the magnon dissipation rate,
as shown in Fig.\,\ref{fig:S3}d, the retrieval
efficiency approaches unity. If operating at low
temperatures, the magnon linewidth can be reduced
to $0.042$ MHz \footnotemark, which will lead to
a saturated efficiency of $0.95$ for a storage
time $\frac{2\pi}{\Delta\omega}=100\
\mathrm{ns}$.\footnotetext{Spencer, E. G.,
LeCraw, R. C., and Linares, R. C., Jr.
\textit{Phys. Rev.} \textbf{123}, 1937 (1961)}

By numerical calculation, we further studied the
memory for different input pulse detuning and
frequency gradient of the magnons. From our
analytical solutions, we can expect the effective
bandwidth of the MGM to be about
$\kappa'_{a,0}=\kappa_{a,0}+\frac{\pi\left|g_{0}\right|^{2}}{\Delta\omega}\approx\frac{\pi\left|g_{0}\right|^{2}}{\Delta\omega}$
for $N\rightarrow\infty$. However, in experiments
we have a finite $N$, and therefore the bandwidth
is also limited by the bandwidth of the gradient
magnon spectrum, which is about $N\Delta\omega$.
Therefore, the highly efficient memory works in a
bandwidth of
$\min\{\frac{\pi\left|g_{0}\right|^{2}}{\Delta\omega},\
N\Delta\omega\}$. From Fig.\,\ref{fig:S4}a, the
bandwidth is about $8\Delta\omega$, which is
consistent with our theory. Figure\,\ref{fig:S4}b
shows that the storage time is inversely
proportional to $\Delta\omega$, which also agrees
with our expectation.

\subsection{Measurement Scheme}

\begin{figure}[htpb]
\begin{centering}
\includegraphics[width=14cm]{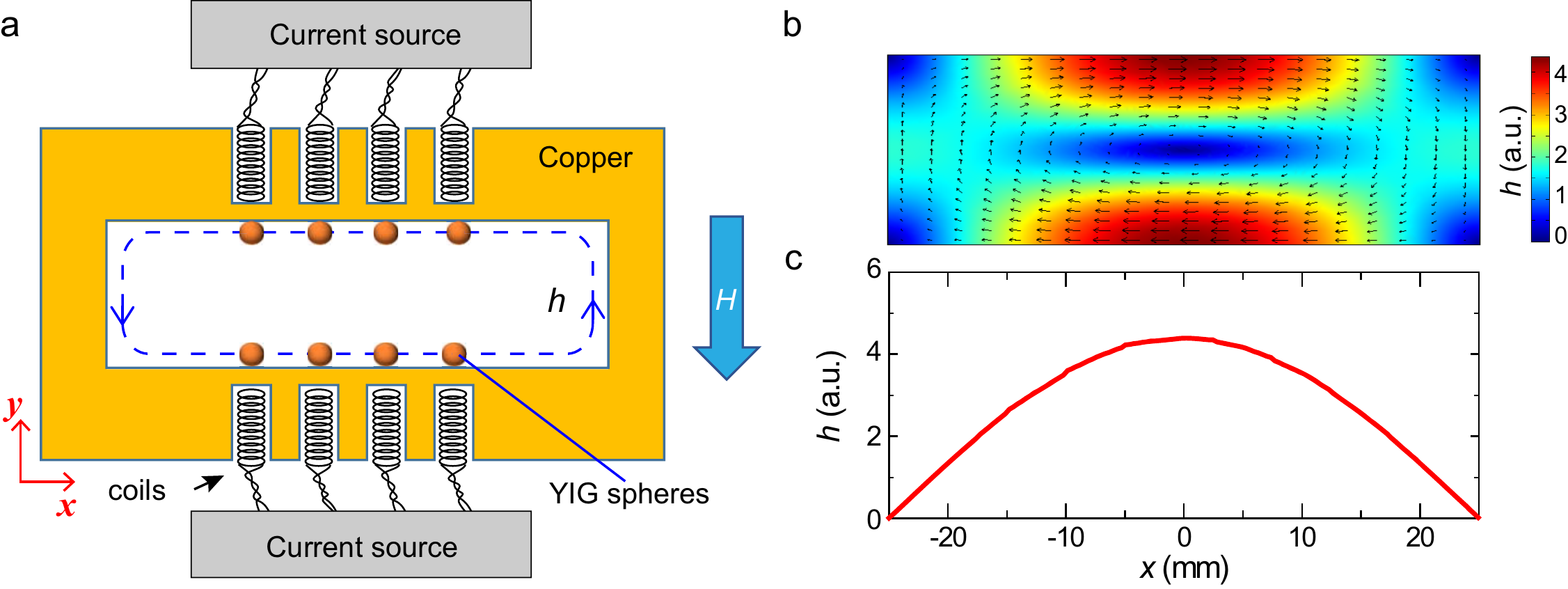}
\par\end{centering}
\caption{a, Device schematic of the MGM
consisting of eight YIG spheres. b, The intensity
and direction of magnetic field distribution of
the cavity TE$_{110}$ mode. Only the $xy$
cross-section is shown because the magnetic field
of this mode is almost uniform along the $z$
direction. c, The cosine function-like magnetic
field distribution along the $x$ direction on the
cavity wall where the YIG spheres are
located.}\label{fig:S5}
\end{figure}

Figure\,\ref{fig:S5}a illustrates the device
schematic of the MGM with eight YIG spheres. Note
that the radius (3 mm) of the small coils is much
larger than the YIG sphere radius to ensure the
magnetic fields generated by the coils are
uniform at the position of the YIG sphere (1 mm
above the coil). As a result, the coils have to
be placed 6.5 mm apart from one another, and so
are the YIG spheres. Since the magnetic field of
the cavity $\mathrm{TE}_{110}$ mode has a cosine
distribution along the $\overrightarrow{x}$
direction inside the cavity (Figs.\,\ref{fig:S5}b
\& c), different YIG spheres experience different
magnetic field strengths. Therefore the coupling
strength of the magnon modes with the cavity mode
are not identical, and this contributes to the
non-ideality of the MGM. Such non-ideality is
unavoidable, but in our experiment efforts such
as placing the YIG spheres as close to the center
as possible have been taken to reduce the
non-ideality in the device.

\vskip 2mm

\begin{figure}[htpb]
\begin{centering}
\includegraphics[width=14cm]{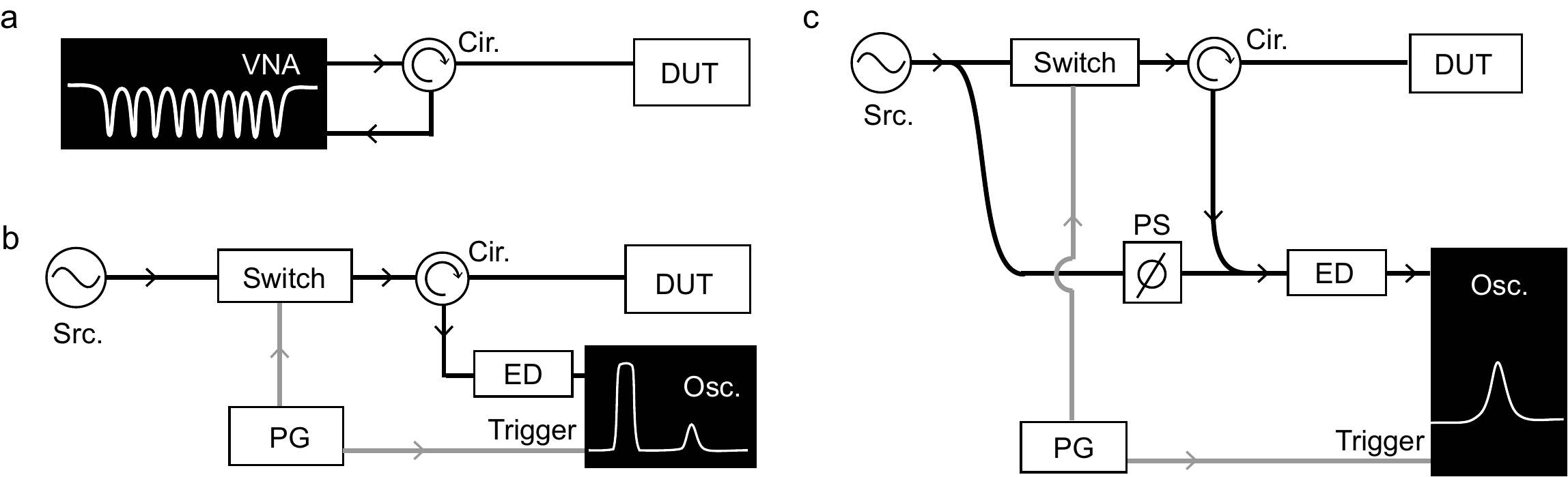}
\par\end{centering}
\caption{Schematic of the measurement setup for
the frequency spectrum of the cavity reflection
(a), the time trace after a pulse excitation (b),
and the coherence of the retrieved pulse (c),
respectively. VNA: vector network analyzer; DUT:
device-under-test; Cir.: circulator; Src.:
microwave source; ED: envelop detector; PG: pulse
generator; Osc.: oscilloscope; PS: phase
shifter.}\label{fig:S6}
\end{figure}

There are three types of measurement involved in
our experiments.

(1) First is the \textit{cavity reflection
spectrum measurement}. The input signal is
provided and the reflected signal is detected by
a vector network analyzer, as indicated in
Fig.\,\ref{fig:S6}a. A circulator is used to
separate the input and reflected signals to avoid
undesired interference.

(2) The second type of measurement is the
\textit{time trace measurement}, which is carried
out using a high-speed oscilloscope
(Fig.\,\ref{fig:S6}b). The signal from a
microwave source is modulated by a pulse
generator through a transistor\textendash
transistor logic (TTL) switch to obtain a pulsed
microwave signal, which is sent into the device,
and then a retrieval pulse can be measured after
the pre-programmed retrieval time.

(3) The third type of measurement is the
\textit{coherence measurement}, as shown in
Fig.\,\ref{fig:S6}c. The scheme is similar to the
time trace measurement but with an interferometer
added. The input signal is split into two
branches, one of which is used as the reference
to interfere with the output signal. The setup is
very similar to the Mach-Zehnder interferometer
commonly used in optical measurements. By varying
the phase of the reference, the amplitude of the
interference signal changes, and their relation
is measured to characterize the retrieval pulse
coherence.

\subsection{Experimental Imperfections}

The operation of the MGM requires the hybrid
magnon-photon modes evenly distributed in the
frequency domain, which will give a perfect
constructive interference at the pre-programmed
retrieval time $T$, and best suppression of the
fringes at other times. Imperfection in the
frequency distribution will result in
deteriorated signal re-construction. This is
clearly illustrated by the comparison given in
Fig.\,\ref{fig:S7}. For a uniformly distributed
spectrum, the retrieval pulse is very clean;
while for the non-uniformly distributed spectrum,
the retrieval pulse is severely distorted.

\begin{center}
\begin{figure*}[htpb]
\begin{centering}
\includegraphics[width=16cm]{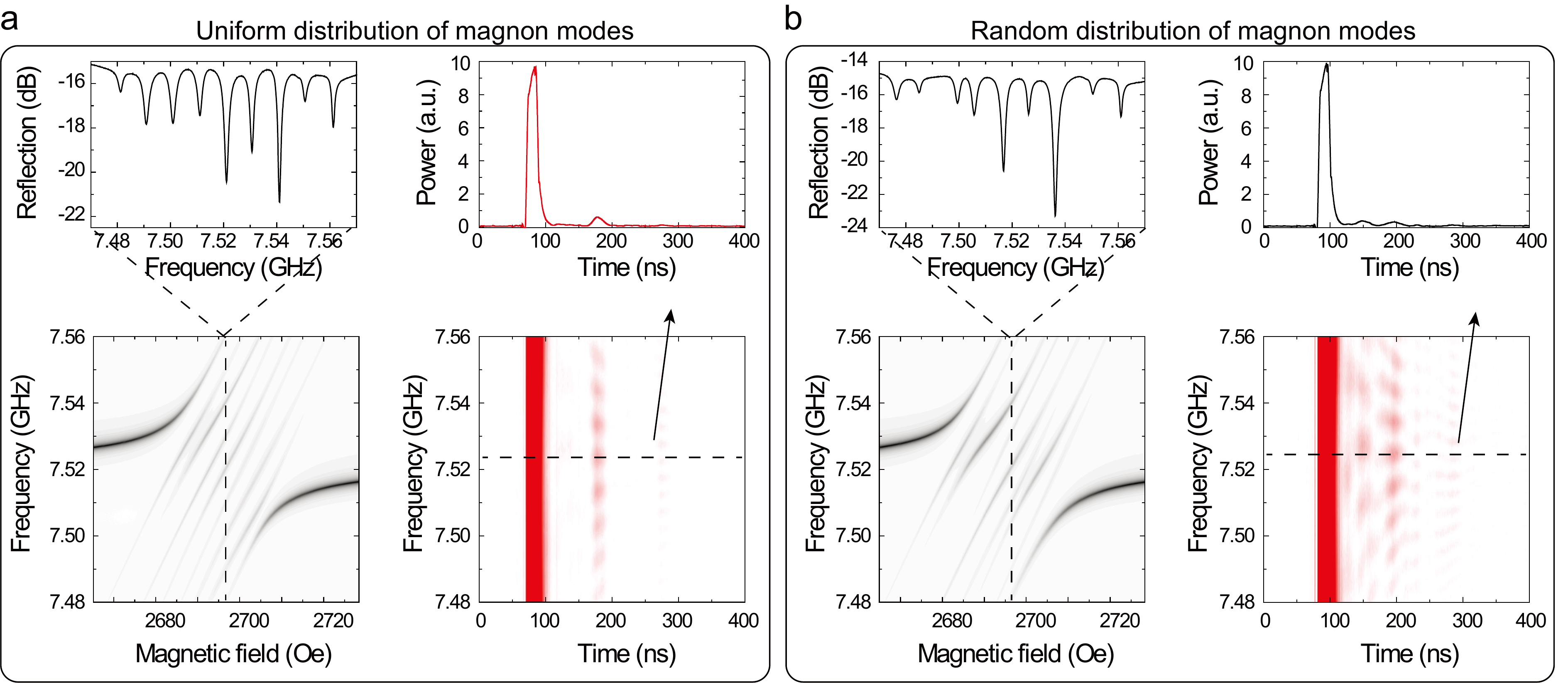}
\par\end{centering}

\protect\protect\protect\caption{Comparison of
the frequency spectrum and time trace for the
uniformly (a) and randomly (b) distributed magnon
mode scenarios, respectively. In each panel, the
left side is the frequency spectrum and the right
side is the time trace; the bottom rows are the
results at different bias magnetic fields, while
the top rows are the specific cases indicated by
the dashed lines in the bottom rows. }
\label{fig:S7}
\end{figure*}

\par\end{center}

In our experiments, we can precisely tune the
small coil to obtain a near perfect gradient for
the magnon modes. The small experimental
imperfections only have very slight effects on
the performance of the MGM. To investigate the
influence of the experimental imperfection on the
efficiency of the MGM, we simulated the
efficiencies for various parameters, with random
perturbations to the ideal case, where
\begin{align}
g_{j} & =g_{0}\times(1+\xi_{j}),\\
\omega_{j} & =\omega_{a}+(j-\frac{N-1}{2})\Delta\omega_{m}+\Delta\omega_{m}\times\xi_{j}.
\end{align}
Here, $\xi_{j}\in[-0.1,0.1]$ is a uniformly distributed random variable,
which means the variations of magnon frequency and coupling strengths
are within a range of $\pm10\%$.

The results for $500$ sets of different
parameters are shown in Fig.\,\ref{fig:S8}, with
the mean value of $\zeta$ being about 0.30, while
the standard deviation is $0.009$. For such a
high imperfection up to $10\%$, the MGM still
shows a very good ability for signal
re-construction. Therefore, the performance of
the MGM is very robust against experimental
imperfections.

\begin{center}
\begin{figure}[htpb]
\begin{centering}
\includegraphics[width=8cm]{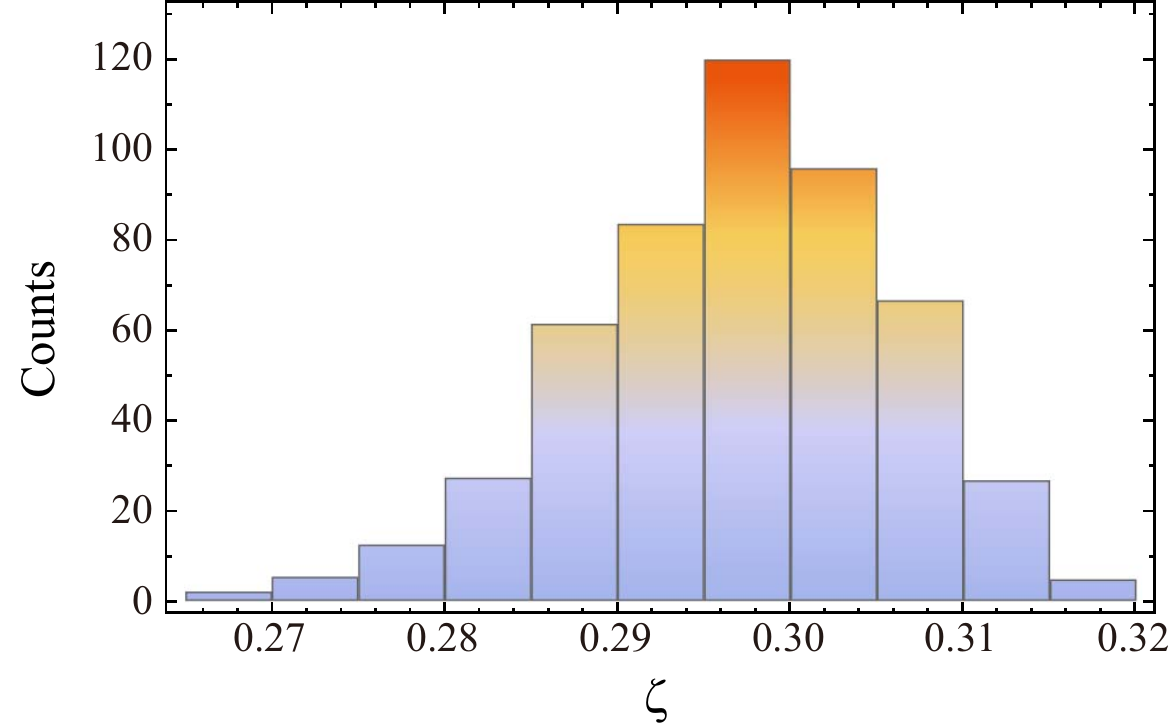}
\par\end{centering}

\protect\protect\protect\caption{The statistics of the numerically solved memory efficiency for randomly
varied parameters $g_{j}$ and $\omega_{j}$. }

\label{fig:S8}
\end{figure}

\par\end{center}
\end{document}